\documentclass[sigconf]{acmart}

\AtBeginDocument{%
  \providecommand\BibTeX{{%
    \normalfont B\kern-0.5em{\scshape i\kern-0.25em b}\kern-0.8em\TeX}}}
    
\copyrightyear{2022} 
\acmYear{2022} 
\setcopyright{rightsretained} 
\acmConference[ICSE '22]{44th International Conference on Software Engineering}{May 21--29, 2022}{Pittsburgh, PA, USA}
\acmBooktitle{44th International Conference on Software Engineering (ICSE '22), May 21--29, 2022, Pittsburgh, PA, USA}
\acmDOI{10.1145/3510003.3510208}
\acmISBN{978-1-4503-9221-1/22/05}

\usepackage{threeparttable}
\usepackage{adjustbox}
\usepackage{color}
\usepackage{hyperref}

\hypersetup{breaklinks=true,
            colorlinks=true,
            citecolor=Blue,
            urlcolor=Blue,
            linkcolor=Blue,
            } % set automatically by hyperref?
\usepackage{xspace}
\newcommand{\sys}{\textit{$\mu$AFL}\xspace}
\newcommand{\E}{\textbf{E}\xspace}
\newcommand{\N}{\textbf{N}\xspace}
\newcommand{\eg}{e.g., }
\newcommand{\ie}{i.e., }
\renewcommand{\paragraph}[1]{\vspace{5pt}\noindent\textbf{#1}}

\newcommand{\tabincell}[2]{\begin{tabular}{@{}#1@{}}#2\end{tabular}}
\newcommand{\BB}{\textit{\textit {LCSAJ\_BB}}\xspace}
\newcommand{\BBs}{\textit{\textit {LCSAJ\_BBs}}\xspace}

\usepackage{blindtext} % list inline
\usepackage[inline]{enumitem}
\setlist[enumerate]{parsep=\smallskipamount}
\usepackage{amsmath} % for etm packet style
\usepackage[ruled,vlined]{algorithm2e}
\usepackage{algpseudocode}
\usepackage{listings}
\usepackage{xcolor}
\usepackage{multirow}

\definecolor{codegreen}{rgb}{0,0.6,0}
\definecolor{codegray}{rgb}{0.5,0.5,0.5}
\definecolor{codepurple}{rgb}{0.58,0,0.82}
\definecolor{backcolour}{rgb}{0.95,0.95,0.92}
\definecolor{textred}{rgb}{1,0,0}
\lstdefinestyle{mystyle}{
    backgroundcolor=\color{white},   
    commentstyle=\color{codegreen},
    keywordstyle=\color{magenta},
    numberstyle=\tiny\color{codegray},
    stringstyle=\color{codepurple},
    basicstyle=\ttfamily\footnotesize,
    breakatwhitespace=false,         
    breaklines=true,                 
    captionpos=b,                    
    keepspaces=true,                 
    numbers=left,
    xleftmargin=1.5em,
    numbersep=5pt,                  
    showspaces=false,                
    showstringspaces=false,
    showtabs=false,                  
    tabsize=2
}

\usepackage[utf8]{inputenc}
\usepackage{tabularx}
\usepackage{xparse}

\usepackage{pifont}% http://ctan.org/pkg/pifont
\newcommand{\cmark}{\ding{51}}%
\newcommand{\xmark}{\ding{55}}%

\NewDocumentCommand{\statcirc}{ O{#2} m }{%
    \begin{tikzpicture}
    \fill[#2] (0,0) circle (1.0ex); % Fill circle with base colour (arg#2)
    \fill[#1] (0,0) -- (180:1ex) arc (180:0:1ex) -- cycle; % Fill a half circle filled with second colour (arg#1), if specified
    \end{tikzpicture}
}

\begin{document}

\title{\sys: Non-intrusive Feedback-driven Fuzzing for Microcontroller Firmware}


% \author{Wenqiang Li$^{\dag,\|}$, Jiameng Shi$^{\ddag}$, Fengjun Li$^{\S}$, Jingqiang Lin$^{\P}$, Wei Wang$^{\dag}$, Le Guan$^{\ddag}$}
% \affiliation{
%  $^{\dag}$ State Key Laboratory of Information Security, Institute of Information Engineering, Chinese Academy of Sciences\\
%  $^{\ddag}$ Department of Computer Science, the University of Georgia, USA\\
%  $^{\S}$ Department of Electrical Engineering and Computer Science, the University of Kansas, USA \\
%  $^{\P}$ School of Cyber Security, University of Science and Technology of China\\
%  $^{\|}$ School of Cyber Security, University of Chinese Academy of Sciences\\
% %
% \{liwenqiang, wangwei\}@iie.ac.cn, \{jiameng, leguan\}@uga.edu, fli@ku.edu, linjq@ustc.edu.cn
% \country{}
% }

\author{Wenqiang Li}

\affiliation{%
  \institution{State Key Laboratory of Information Security, Institute of Information Engineering, Chinese Academy of Sciences}
  \institution{School of Cyber Security, UCAS}
  \state{Beijing}
  \country{China}
}
\email{liwenqiang@iie.ac.cn}

\author{Jiameng Shi}
\affiliation{%
  \institution{Department of Computer Science, \\ the University of Georgia}
  \city{Athens}
  \state{Georgia}
  \country{USA}
}
\email{jiameng@uga.edu}

\author{Fengjun Li}
\affiliation{%
  \institution{Department of Electrical Engineering and Computer Science, \\ the University of Kansas}
  \city{Lawrence}
  \state{Kansas}
  \country{USA}
}
\email{fli@ku.edu}

\author{Jingqiang Lin}
\affiliation{%
  \institution{School of Cyber Security, University of Science and Technology of China}
  \city{Hefei}
  \state{Anhui}
  \country{China}
}
\email{linjq@ustc.edu.cn}

\author{Wei Wang}
\affiliation{%
  \institution{State Key Laboratory of Information Security, Institute of Information Engineering, Chinese Academy of Sciences}
  \state{Beijing}
  \country{China}
}
\email{wangwei@iie.ac.cn}

\author{Le Guan}
\affiliation{%
  \institution{Department of Computer Science, \\ the University of Georgia}
  \city{Athens}
  \state{Georgia}
  \country{USA}
}
\email{leguan@uga.edu}

\begin{abstract}
Fuzzing is one of the most effective approaches
to finding software flaws.
% automatically by taking advantage of
% abundant testcases with coverage-guided mutation mechanism.
However, applying it to microcontroller firmware incurs
many challenges.
% First, firmwares lack important runtime support (\eg~fuzzing instance automation, execution information collection) to facilitate efficient fuzzing, making it difficult to conduct fuzzing on real devices.
% Second, it is not scalable to rehost firmware that contains
% peripheral-specific or proprietary code.
For example, rehosting-based solutions cannot accurately
model peripheral behaviors and thus cannot be used to fuzz the
corresponding driver code.
% tremendous
% difference in the run-time environment on 
% microcontroller systems and commodity computing systems, such as PCs.
% For example, the microcontroller
% firmware does not have a signal mechanism to report memory access
% violations, which are critical in fuzzing.
% As another example, the efficient fork server
% implemented in AFL relies on the process-level fork
% mechanism, which is not in place in microcontroller-based systems.
% This makes it extremely difficult to test peripheral drivers
% or proprietary code which cannot be easily rehosted.
% Existing works focus on the emulator based solution or rehosting.
% However,these solutions
% could not run and cover the whole firmware,
% especially the peripheral driver layer,
% which is the communication channel with the outside world
% and the typical attack point of IoT devices.
% Besides, some commercial libraries are
% arduous to disassemble and instrument.
% Consequently, how to fuzz the whole firmware,
% in particular driver libraries, 
% is still an open challenge for the microcontroller.
In this work, we present \sys, a hardware-in-the-loop approach
to fuzzing microcontroller firmware.
It leverages debugging tools in existing embedded system development to construct an AFL-compatible fuzzing framework.
Specifically, we use the debug dongle to
bridge the fuzzing environment on the PC and the
target firmware on the microcontroller device.
To collect code coverage information without costly code instrumentation,
\sys relies on the ARM ETM hardware debugging feature,
which transparently collects the instruction trace and
streams the results to the PC.
% to collect the path coverage information
% to guide the fuzzing process.
% In this way, we can fuzz the whole firmware,
% even without source code.
However, the raw ETM data is obscure and
needs enormous computing resources to recover the actual instruction flow.
We therefore propose an alternative representation of code coverage,
which retains the same path sensitivity as the original
AFL algorithm,
% called dynamic basic block (\BB)
but can directly work on the raw ETM data without matching them
with disassembled instructions.
To further reduce the workload,
we use the DWT hardware feature to selectively
collect runtime information of interest.
We evaluated \sys on two real evaluation boards from two major vendors: NXP and STMicroelectronics.
% equipped with ETM and DWT.
% from different vendors.
% including TWR-K64F120M from NXP, STM32H7B3I-EVAL from STM and 
% SAM E54 Xplained Pro from Microchip.
% Our evaluation result shows that \sys
% could be widely applied to all the microcontroller 
% equipped with ETM and DWT,
% and recover the whole control-flow of the firmware perfectly.
With our prototype, we discovered ten zero-day bugs in
the driver code shipped with the SDK of STMicroelectronics
and three zero-day bugs in the SDK of NXP.
Eight CVEs have been allocated for them.
Considering the wide adoption of vendor SDKs in real products,
our results are alarming. 

\end{abstract}





\maketitle

\section{Introduction}
\label{sec:intro}

% & \tabincell{c}{\textbf{Fuzzing Performance} \\ \textbf{(exec/s))}} 

Internet-of-Things (IoT) has become an integral part
of our digital lives.
% is more than 10 billion by 2021~\cite{iottotalnumber},
% indicating that everybody has an average of more than one
% IoT device.
% In the meantime, the amount of connected IoT devices is growing 
% at a rapid rate over the past five years~\cite{iotgrowrapidly}
% and a faster pace is expected~\cite{iotgrowrate}.
For example, many people install smart thermostats to remotely control the temperature and humidity of their homes.
Smart alarm systems are used to monitor home and workplace and raise alarms when detecting burglars. Fitness trackers and smart health bands are also widely used to continuously monitor personal health data such as heart rate and blood oxygen level.
The key component of many IoT devices is the microcontroller unit (MCU), which is a tiny, custom-built, and cost-efficient system-on-chip (SoC)~\cite{mcumarket}. 

The rapid evolution of the MCU ecosystem, on the one hand, has made our lives easier and more convenient than ever before, on the other hand, it also introduces a large number of vulnerable MCU products in the wild.
%, partially due to the pressure of time-to-market. 
For example, several high-profile vulnerabilities have been reported recently for the ESP8266 and ESP32 communication and WiFi co-processors, which have been adopted in millions of IoT devices. These vulnerabilities (e.g., {\em EAP client crash}, {\em zero PMK installation} and {\em beacon frame crash}~\cite{wifiesp}) allow the adversaries to hijack or crash the session
of ESP32/ESP8266 products. Meanwhile, vulnerabilities of the FreeRTOS TCP/IP stack can be used to launch remote code execution
and steal private information
~\cite{freertostcpipbugsaffect,freertostcpipbugs}. More recently, BadAlloc~\cite{BadAlloc} leaves a large number of IoT devices
exposed to adversaries. The compromise of an MCU product may lead to serious consequences such as privacy leakage, financial loss, or even human injury and death. To prevent or mitigate such attacks, extensive security testing during the development phase is imperative.

% Unfortunately, as mentioned before, other than manually
% reviewing the code,
% MCU developers often do not feel up to find bugs
% due to a lack of sophisticated tools.
% % find they lack a sophisticated tool that can help them find bugs.
% Specifically, existing dynamic analysis solutions either only 
% target hardware-independent part of firmware~\cite{HALucinator,rehosting},
% or rely on inaccurate emulation~\cite{automatedrehosting,P2IM,uemu},
% or require non-trivial 
% human involvement and domain knowledge~\cite{avatar2,avatar,ruge2020frankenstein,garbelini2020sweyntooth}.
% Notably, all existing works cannot be applied to test peripheral drivers,
% which are indispensable for MCU firmware to manage
% I/O interfaces, but are error-prone compared with other code.

\begin{table*}[t]
\caption{Comparison with the state-of-the-art solutions}
\begin{center}
% \begin{adjustbox}{max width=\textwidth}
\resizebox{\textwidth}{!}{
\begin{tabular}{l|c|c|c|c|c|c}
\textbf{Basic Method} & \textbf{Solutions} & \tabincell{c}{\textbf{Hardware} \\ \textbf{Independent Code}} & \tabincell{c}{\textbf{Driver} \\ \textbf{Code}} & \tabincell{c}{\textbf{Support} \\ \textbf{Fuzzing}} & \tabincell{c}{\textbf{Require} \\ \textbf{Source Code}} & \tabincell{c}{\textbf{Require} \\ \textbf{Hardware}} \\
\hline
Rehosting 	& HALucinator~\cite{HALucinator}, {P$^{2}$IM}~\cite{P2IM}, $\mu$Emu~\cite{uemu}, etc. 	& \cmark & \xmark 	& \cmark &  N &  N	\\
Porting & Para-rehosting~\cite{rehosting}    	& \cmark & \xmark 	& \cmark &  Y &  N	\\
Forwarding Hardware Interactions & Avatar~\cite{avatar}, Avatar$^2$~\cite{avatar2}, SURROGATES~\cite{SURROGATES}, Inception~\cite{corteggiani2018inception}, etc.  	& \cmark & \xmark$^{*}$ 	& \xmark$^{*}$ & N &  Y	\\
\textbf{Fully On-device Execution}	& \textbf{\sys (proposed)} 	& \cmark & \cmark 	& \cmark & N  &  Y\\
\end{tabular}
}
\label{tab:comparison_of_systems}
% \end{adjustbox}
\end{center}
\flushleft
\scriptsize{
*: Theoretically, these solutions support fuzzing the driver code.
However, the significant overhead on state syncing renders
fuzzing driver code impractical.
Existing solutions leverage real devices
to boot the firmware in QEMU to a state where analyzing hardware-independent code is possible.
}
\end{table*}

%We have not correspondingly developed our capability to test the security of MCU products.
However, existing firmware security testing approaches demonstrate their own limitations when applied to embedded firmware testing (see Table~\ref{tab:comparison_of_systems} for an overview of the issues).
For example, the emulation-based rehosting technique with application to firmware analysis has been extensively studied in recent years~\cite{HALucinator,P2IM,dice,uemu,jetset,PRETENDER,laelaps}, but accurately modeling the behavior of diverse peripherals remains the main research challenge. Several approaches such as {P$^{2}$IM}~\cite{P2IM}, DICE~\cite{dice}, Laelaps~\cite{laelaps}, PRETENDER~\cite{PRETENDER}, $\mu$Emu~\cite{uemu}, and Jetset~\cite{jetset} propose to learn an approximate peripheral model using the symbolic execution, access-pattern matching, and machine learning techniques. However, the learned models are inaccurate in general. With inaccurate models, such rehosting approaches cannot boot firmware with complex peripherals such as USB. When the execution trace is not exactly the same as that on the real device, even though the firmware can be ``successfully'' booted, it is inadequate for many security analysis tasks. 
The emulation-based rehosting technique is typically used to test \emph{hardware-independent code} after the target firmware has passed through the booting process. 

Another direction is to leverage the hardware abstraction layer (HAL) available in MCU firmware to avoid modeling peripherals. For example, HALucinator~\cite{HALucinator} automatically detects
HAL libraries and replaces them with host implementations. Para-rehosting~\cite{rehosting} provides common HAL backend implementations to help port MCU firmware to native hosts. However, both approaches cannot test the peripheral driver which never runs.

Finally, some hardware-in-the-loop (HITL) solutions run the firmware in the QEMU emulator while forwarding peripheral I/O operations to real devices~\cite{avatar,avatar2,SURROGATES,corteggiani2018inception}. Although high fidelity is preserved, these HITL-based approaches require frequent and expensive switching (and syncing) between QEMU and hardware, incurring significant performance overhead. For example, the emulation speed is in the order of tens of instructions per second when frequent hardware interaction is needed~\cite{avatar}. As a result, these approaches 
are typically used to analyze \emph{hardware independent code} after the firmware is fully booted with the help of the forwarding mechanism. To the best of our knowledge, none of the existing HITL work supports firmware fuzzing. 

In this work, we propose a new fuzzing solution called {\em micro-AFL} (\sys), which is specifically designed for ARM-based MCU devices, to assist developers in locating potential software bugs in the firmware. While our approach can cover the entire software stack, in this paper we focus on its use in testing the low-layer code such as peripheral drivers, which we believe is not properly supported by the existing work. By running the target firmware directly on the target device, our approach supports full-stack testing with high fidelity. \sys~requires developers to have access to the prototype development boards and JTAG- or SWD-based debug dongles, which are essential hardware tools available and used in virtually every embedded system development environment.

\sys is designed to be modularized and extensible so that existing mature fuzzers can be directly integrated (e.g., we adopt AFL~\cite{afl} in our prototype implementation but any other fuzzers can be used as a drop-in replacement). To achieve this goal, {\bf \em{(i)}} \sys~decouples the \emph{execution engine} from the rest of a fuzzer (collectively called \emph{fuzzing manager} in this paper). 
% which allows it to  call the rest \emph{fuzzing manager} to perform testcase generation, feedback analysis, crash detection, etc.
Specifically, we run the target firmware on the real hardware,
and run the fuzzing manager on the PC that coordinates the fuzzing process with the help of the debug dongle.
The execution information is also streamed to the fuzzing manager where the analysis is conducted. 
In this way, we keep the fuzzing manager on the PC agnostic to the execution engine.
% our approach also relies on a powerful PC that runs AFL~\cite{afl} to generate testcases and
% coordinate the fuzzing instances on the real MCU device (thus the name \sys).
% In essence, we separate the execution engine from the original AFL
% and move it to the development board for high fidelity firmware execution.
{\bf \em{(ii)}}  To enable communication between the development board and the PC, we re-purpose the debug dongles, which have the highest level of control to the board. Therefore, \sys~can efficiently and effectively feed the testcases over the board, pull the run-time execution status, and start/suspend/stop the target. {\bf \em{(iii)}}  Finally, to collect code coverage information of each testcase, which is essential for grey-box feedback-driven fuzzers, \sys~leverages a hardware feature called Embedded Trace Macrocell (ETM)~\cite{etmv3} that transparently generates the instruction trace.
The trace is streamed directly to the PC via five additional pinouts (four for data and one for clock).
Although ETM incurs additional cost, we adopt it in this work
for two reasons. First, it enables transparent trace collection. In other words, no code instrumentation is required.
This feature makes our work free from rewriting the binary, since most of the third-party libraries are distributed
as stripped binaries
% or hex form (\eg~SoftDevice in Nordic~\cite{softdevice})
and no robust binary rewriting tool is available to facilitate instrumentation-based trace collection.
Second, we argue that prototype development boards with ETM pinouts are only needed in the development phase.
After the firmware has been fully tested, the released products do not need to be equipped with these features. 
This small investment at the development stage yields a good return on investment (ROI) for manufacturers in the long term,
considering the expensive recalls that may happen later.

\sys features two key components:
\textit{online trace collector}
and \textit{offline trace analyzer}.
They collect ETM data on the device and
parse the results on the PC respectively.
% We called them \textit{online-trace-collector}
% and \textit{offline-trace-analyzer} respectively.
% Based on this observation, we first propose \sys,
% a hardware-supported fuzzing platform 
% committed to MCU peripheral driver vulnerabilities discovery.
% \sys builds on the mature fuzzing tool AFL
% and contains two core components:
% \begin{enumerate*}[label={\alph*)},itemjoin={{, }}, itemjoin*={{, and }}, font={\color{red!50!black}\bfseries}]
% \item Online-Trace-Collector
% \item Offline-Trace-Analyser.
% \end{enumerate*}
% In the Online-Trace-Collector component,
When a testcase is available on PC (generated by AFL),
\sys sends it into a reserved memory on the board
via the debug dongle.
% and starts this execution instance from the entry point of the firmware.
% while the firmware running from \textit{Reset\_Handler}
% to the fuzzing start point.
% This concurrent operation contributes to improving the performance.
% At the point where the testcase is consumed
% for the first time,
At the point where the testcase is consumed for the first time,
the online trace collector activates ETM to collect
the instruction trace,
% In the meantime, the generated ETM packets are
and streams the data to the PC.
While collecting the ETM stream,
the online trace collector also applies configurable filters
via the Data Watchpoint and Trace (DWT) unit~\cite{m4_tech_ref} to suppress unnecessary ETM packet generation.
This not only reduces the amount of
tracing data for transmission, but also
avoids analyzing useless packets on PC.
% Specifically,
% instruction trace filters can help the online trace collector to
% focus on a code snippet of interest.
% and improve transfer performance.
% This is also beneficial to the offline analysis performance.

% or start or stop working when a specific variable value hit.
% This is significant and helpful, 
% especially when we are only interested 
% in the peripherals driver layer.

The offline trace analyzer runs on the PC
and processes the raw ETM data.
The result is provided to AFL to maintain the bitmap of code coverage.
Decoding the raw ETM data to get the branch information is expensive
since it needs to disassemble the firmware and align
the instructions with the raw trace~\cite{ge2017griffin} (see \textbf{RQ1} in Section~\ref{sec:evaluation}).
We address this problem by using
a kind of special basic block generated at runtime.
This allows us to directly use the raw ETM data without disassembling,
but still retains path-sensitivity needed to calculate code coverage.
% With this design, we do not need to disassemble
% the target firmware to recover the instruction trace,
% but still capture new branch edges,
% which is fundamental for AFL to decide
% interesting testcases.
The offline trace analyzer also uses a software based approach
to filter out uninteresting ETM packets that cannot be filtered
by the online trace collector.
% Finally, we design a hash algorithm suitable
% for recording the branch coverage on top of \BB.
% The resulting bitmap is compatible
% with other components in AFL,
% allowing \sys to enjoy the intelligence of the
% state-of-the-art fuzzer with minimal modifications.

% \sys only depends on the raw ETM packet
% to identify each execution path
% uniquely without disassembly.
% \sys combines the execution information 
% including branch target, sequential instruction number, 
% and exception entry and exit point,
% which is the original information encoded in the ETM packets,
% to recover and represent each round execution.
% Without disassembly, \sys could avoid the issues
% introduced by the code obfuscation, code compression and so on.
% \sys also provides an offline filter 
% to compensate for the restricted capacity
% of the online filters
% and to address the non-deterministic interruptions issue.
% In addition, \sys design a hash algorithm 
% to tackle the bitmap collision due to the heterogeneous 
% and compact memory address space.

We have implemented a prototype of \sys 
using the SEGGER MCU debugging solution~\cite{seggersdk}.
Then, we used the prototype to test SDKs from two major MCU chip vendors, i.e.,
NXP Semiconductors~\cite{nxp} and STMicroelectronics~\cite{stmicro}.
% We have evaluated multiple peripheral drivers and closed source libraries
% including USB, SD Card, Ethernet, MMCAU, UART, SPI, DAC, ADC, etc.
In particular, we used the USB driver fuzzing as a case study in our evaluation.
At the time of writing, we have uncovered {\em 13 bugs residing in the USB drivers that were not known previously}.
All of them have been confirmed by the vendors
and the patches have been released or scheduled with the newest SDK releases.
% Among them, \gl{XX} have been patched
% and \gl{XX} are pending to be fixed.

In summary, our contributions are three-fold:
% \gl{mention online filters, dynBB, offline interrupt filter, hash function as required by the *significance of your research contributions*}
\begin{itemize}
    \item We propose \sys, the first fuzzing tool that is applicable to the driver code of MCU firmware.
    \sys decouples the execution engine from the fuzzing manager so that existing fuzzing tools can be easily integrated. 
    % Our solution mainly serves MCU developers,
    % who have access to the debug dongle
    % available in the existing development environment.
    % explored ARM ETM and proposed \sys, an automatic
    % non-intrusive feedback-driven fuzzing platform
    % committed to uncovering the peripheral drivers and
    % closed-source libraries vulnerabilities.
    \item We propose using ARM ETM for non-intrusive feedback collection. To improve performance, \sys adopts
    Linear Code Sequence And Jump (LCSAJ) analysis
    to directly process raw ETM data without expensive disassembling.
    % and also leverages ARM DWT to filter out irrelevant ETM packets.
    
    % \item We designed and implemented the three most important filters
    % depending on restricted resources 
    % to resolve the non-deterministic issue
    % and improve the performance.
    
    \item We have implemented and evaluated
    our prototype against two SDKs from major
    MCU chip vendors.
    We used the USB driver as a case study to show how our
    prototype can fuzz real-world driver code.
    The tool has helped us find 13 previously unknown bugs with 8 CVEs allocations. 
\end{itemize}

The source code and the firmware samples used
in the evaluation are available at \url{https://github.com/MCUSec/microAFL}
for future research on this topic.

% The source code and the firmware samples that we have tested
% are provided anonymously at \url{https://anonymous.4open.science/r/ETMFuzz-B104}
% for review purposes. We will open source our tools for future research on this topic.

% For future research on this topic, we 
% will open source our tool (URL omitted due to
% double-blind review restrictions).

% have open-sourced \sys prototype
% at \textcolor{textred}{\url{https://github.com/MCUSec/microAFL}}
\section{Background}
% \subsection{Coverage-guided Fuzzing}
\subsection{American Fuzzy Lop (AFL)}
\label{bg_afl}
Fuzz testing is an automated testing technique used to discover coding errors and security vulnerabilities in software.
It involves inputting abnormal testcases to the software-under-test
in an attempt to make it crash.
American Fuzzy Lop (AFL)~\cite{aflwhitepaper} 
is one of the most successful fuzzing tools.
We roughly split it
into two main components for easy presentation:
an execution engine and a fuzzing manager.
While the former is responsible for running a testcase with the target program,
the latter is responsible for generating new testcases by mutation
based on a genetic algorithm, coordinating the execution, analyzing the
execution information, etc.
Concretely,
AFL first instruments the target program so when the program is executed,
the branch information can be generated and recorded.
The fuzzing manager then forks a new process as the execution engine 
to run the program with the current testcase.
% and initializes a
% identifies new paths by instrumenting random numbers to each basic block at the source code level and mapping them into the bitmap while fuzzing.
% At the beginning, 
% the fuzzing manager creates a new process as the execution engine and initialize a forkserver
% to load the instrumented program.
During execution, the instrumented target program
consumes the testcase and
records the collected branch coverage information into a local bitmap.
% between the fuzzing manager and itself.
The fuzzing manager also aggregates all the local bitmaps into a global bitmap,
and compares the newly generated local bitmap to the global one 
to decide if a new path has been discovered.
A testcase that can increase branch coverage is considered interesting and
will be used in the genetic algorithm to calculate the subsequent testcases.
To report a bug,
the fuzzing manager monitors the execution status of the target program
and leverages crash information as indicators.

% Due to the separated logic between the execution engine
% and the fuzzing manager,
% in \sys, we use the real device as the execution engine 
% and run the fuzzing manager on a PC.
% To let them exchange information (\eg~bitmap),
% \sys uses the debug dongle to bridge the device and the PC.

% With automatically feeding testcases generated by the guided paths,
% AFL could cover more codes and trigger more bugs
% in a short period.
% To obtain the path information of each round execution,
% AFL instruments the unique random number
% at the beginning
% of each basic block at the compile time.
% While running, the random numbers are calculated
% as the index of the bitmap
% to indicate if the new path is discovered.
% Unfortunately, AFL is intended for user-space applications.
% To extend AFL ability to the kernel level, 
% Syzkaller~\cite{syzkaller} enables kernel option \textit{KCOV}
% to collect the execution information.
% Notwithstanding, all these non-black-box fuzzing solutions
% need the source code to instrument the path collection logic.
% To overcome this issue,
% kAFL~\cite{schumilo2017kafl} collects the control flow
% information by QEMU-PT which is based on the hardware component.
% However, all these solutions are meant for the x86 platform,
% but the ARM Cortex-M platform.
% FaceDancer~\cite{facedancer} is a fuzzing tool 
% that could fuzz the USB driver of the ARM Cortex-M boards,
% but in a dummy way which is not efficient.
% Based on our knowledge, there is still no
% solution that could fuzz ARM Cortex-M board drivers
% in a non-intrusive feedback-driven way.

\subsection{Analysis of MCU Firmware}
\label{sec:mcuandfirmware}
MCU is a special-purpose System-on-Chip
that cares about real-time processing capability,
low power consumption and costs.
% Unlike processors shipped 
% with high-end systems,
% MCUs work at a lower frequency for
% low power consumption 
% and the average price is under \$1.
They are widely used in different application fields,
such as wearable, smart home, industrial automation, etc.
The execution environment of MCU firmware is significantly
different from the traditional OSs, making many existing
binary analysis tools including AFL inapplicable.

Unlike traditional software which assumes an OS layer that provides
an abstract view of hardware, MCU firmware
runs on bare metal or only includes an OS library (\eg~RTOS) for simple multi-task management.
Therefore, it compiles the driver code of peripherals and the application code together
to form a single-address-space program.
The peripheral I/O operation is performed by accessing
the memory-mapped registers.
Due to the diversity of peripherals, 
dynamic analysis of MCU firmware is extremely challenging.
Although the rehosting technique has made some breakthroughs
to test the hardware-independent part of the 
firmware~\cite{HALucinator,P2IM,dice,uemu,jetset,PRETENDER,laelaps,rehosting},
no existing work can test the driver code.

\subsection{Hardware-Supported Instruction Trace Collection}
\label{sec:trace_bg}
Program instruction trace is helpful in many program analysis
applications, such as
performance profiling~\cite{milenkovic2008performance,bhansali2006framework}, 
fuzz testing~\cite{chen2019ptrix}, 
control flow integrity enforcement~\cite{ge2017griffin},
root cause analysis~\cite{REPT, du2020hart},
debugging~\cite{ning2017ninja}, etc. 
Compared with software instrumentation,
modern processors support capturing 
the instruction traces by
hardware components
to reduce the overhead.
For instance, 
Intel incorporates its hardware instruction trace
feature, known as Processor Trace or PT~\cite{sdm}
to all its Core processors starting from Broadwell.
The counterpart of ARM is called
\textit{Embedded Trace Macrocell} (ETM)~\cite{m4trm} or
\textit{Program Trace Macrocell} (PTM)~\cite{lee2017using}\footnote{ETM and PTM are
similar techniques for different Arm processor lines. 
We use ETM to refer to both in this paper.}. 
These implementations are quite similar to each
other. 
Both are designed to efficiently rebuild the whole instruction trace 
assuming that the corresponding machine code is available. 
More specifically, a dedicated hardware component
emits a stream of control flow packets. 
Then a decoder is used to reconstruct a unique execution path 
by matching the control flow data to the disassembled machine code.
% In this work, we leverage the hardware-generated
% instruction trace to construct the branch coverage 
% used for feedback-driven fuzzing.

% The Intel PT and its applicability to numerous problems 
% have been thoroughly investigated in earlier 
% literature~\cite{ge2017griffin, kwon2018vm, thalheim2016inspector}.
% ARM ETM receives minimal attention compared with Intel PT, 
% partly because of its lack of documentation
% \footnote{ARM documentation
% is obscure to system developers 
% because the target readers are hardware
% engineers that focus on system integration. 
% Intel developer manual is more
% friendly for system software developers.}. 
% We reviewed the ETM behavior and compared
% the main differences between it and Intel PT. 
% We concluded that the subtle
% design differences between them exhibit
% both opportunities as well as
% challenges for us.

\subsubsection{Trace Collecting} 
% Intel PT supports using the DRAM as
% the source for storing the emitted trace data. 
% The OS needs to prepare a number of
% non-contiguous buffers for trace storage. 
% It also supports generating
% an interrupt when any buffer is nearly full. 
% If this happens, the processor must stop to take care of the gathered trace. 
ARM MCUs can optionally implement
a buffer for trace storage on the chip,
termed \textit{Embedded Trace Buffer} (ETB). 
However, based on our study,
ETB is rarely supported on real chips.
Alternatively, ARM also supports
streaming the trace data to an external debugger 
via a physical parallel port,
called Cortex Debug+ETM connector~\cite{cortex_debug_connectors}. 
This is the solution used by \sys.
% To utilize this function, 
% the debugger (or the target itself) 
% needs to configure the ETM as
% well as the \textit{Trace Port Interface Unit} (TPIU). 
% Then, 
% the external debugger (\eg~a PC program) 
% can decode and analyze the trace on the fly or store it 
% for postmortem analysis. 
% It is worth noting that the trace generation 
% and collection are
% non-intrusive to the firmware execution 
% and thus incur no performance overhead.
% Intel documents a similar mechanism called  
% \textit{trace transport subsystem}.
% However, it is platform-specific and 
% we are unaware of any real-world implementations.

\subsubsection{Instruction Flow Reconstruction}
\label{sec:bg:etm}
To reconstruct the execution flow, 
a decoder is needed to
interpret the trace packets and 
align them with the disassembled instructions.
Control flow packets carry information about
\begin{enumerate*}[label={\alph*)},itemjoin={{, }}, itemjoin*={{, and }}]
\item whether a conditional branch is taken or not
\item the target of an indirect branch
\item asynchronous events such as exceptions.
\end{enumerate*}

\paragraph{Conditional Branches.}
% Intel PT uses a bit in a \textit{Taken-Not-Taken} (TNT) packet to 
% indicate if a branching instruction
% is taken or not. 
% Therefore, executing a contiguous code region (basic
% block) only generates one bit at the end which is a branch instruction. 
% Different from this, 
ARM ETM uses one bit in the \textit{P-header} packet to encode whether
the condition of an instruction is true (encoded as \E or 1) 
or false (encoded as \N or 0). 
True means the corresponding instruction is executed.
% Therefore, a bit is generated for every instruction. 
The reason for this design is that in ARM, 
almost all instructions can be
conditionally executed.
% based on the condition flags 
% appended to the instruction.
Taking the \texttt{addeq} instruction as an example, the
\texttt{add} operation is conducted only if the \texttt{Z} flag is set. 
When this instruction is a branching instruction, 
such as \texttt{beq}, 
\E indicates that the branch is taken 
whereas \N indicates the branch is not taken.

\paragraph{Indirect Branches.}
The indirect branch includes indirect calls and
function returns. 
Since the target of an indirect branch 
can only be determined at run time, 
ARM ETM emits a packet containing
the target address when an indirect branch happens.
% in additional to the \textit{P-header} data.
% instruction reference. 
% They also perform some compression to 
% reduce the data generated. 
Such information is encoded into 
% a \textit{Target IP} (TIP) packet in Intel PT 
% and 
a \textit{branch packet}.
% in ARM ETM.

\paragraph{Asynchronous Events.} 
An asynchronous exception could change the
control flow at any execution point. 
% For such events, additional data are emitted.
% Intel PT encodes an exception 
% as a \textit{Flow Update Packet} (FUP)
% followed by a TIP, 
% giving the source and destination 
% of the asynchronous branch, respectively.
Since the current execution location can
already be recovered by the \textit{P-headers}, 
the branch source information is unneeded. 
In particular, the branch source is calculated 
by adding the length of executed instructions from the last branch (determined by \textit{P-headers}) 
to the base address of the last branch target (determined by the previous \textit{branch packet}).
ETM encodes asynchronous events using existing branch packets,
but extends them with supplementary information.
% about the exception by extending existing branch packets. 
For example, it can indicate whether this
branch is caused by an exception rather than a normal call instruction. 
It also indicates the corresponding exception number if this is an exception.
Moreover, ARM MCUs re-purpose
existing instructions for exception returns.
Put simply, if an instruction results in a control flow transfer of
a set of predefined values (\texttt{EXC\_RETURN}),
then this is treated as a return from exception 
and the hardware is responsible for fetching 
the correct target instruction pointer from the exception stack. 
ETM further emits a \textit{return from exception} 
packet to encode such an event. 
With this mechanism, exception entries and returns can be
properly paired.

\paragraph{Direct Branches.} 
With the aforementioned information, 
the decoder can already recover the whole execution flow 
by aligning the trace data with
the disassembled instructions. 
Note that the trace information about the
direct branches is not needed, 
since the target of a direct branch could
be determined by checking the corresponding branch instruction. 
However, ETM optionally supports 
emitting branch packets for direct branches, 
making it easier to 
recover the instruction flow at direct branches.
% without 
% having to refer to the executed code.

\subsubsection{Trace Filtering}
\label{sec:dwt}
It is generally unnecessary to collect the entire instruction trace over time
because 
an analyst might be only interested in a particular code region.
% On the one hand, the amount of trace
% data might overwhelm the buffer quickly. 
% On the other hand, an analyst might
% be only interested in a particular code region.
Trace filtering allows for suspending
trace collection under certain conditions.
% Intel PT facilitates rich filtering mechanisms. 
% It can set filtering rules 
% based on the instruction pointer range, 
% current privilege level (CPL), 
% and control register 3 (CR3) 
% which uniquely identifies a process in major OSs.
% Therefore, an analyst can configure PT 
% to only emit the trace packets for 
% a particular process in user/kernel space 
% for certain code ranges.
ARM ETM supports event-based filtering. 
It defines a set of \textit{ETM event resources} 
that become active when the corresponding event occurs.
These events can be configured by different
comparators provided by the hardware.
% address comparators, data value comparators,
% EmbeddedICE watchpoint comparators\footnote{
% The EmbeddedICE macrocell module
% is the legacy name for the ARM integrated on-chip debugging module.}, 
% context ID\footnote{On ARM processors, 
% the context ID uniquely identifies a process.} comparators, etc. 
When there is a match, the corresponding event becomes active.
For example, when the instruction pointer 
matches the value in an address comparator, 
the corresponding event resource is active.
% When a data value being read/written matches the value 
% in a data value comparator, 
% the corresponding event resource is active. 
% These events can further be combined
% using a boolean operation, generating a complex event. 

The trace generation is controlled in three ways. 
First, when an event is active, 
it can directly enable tracing. 
When it is inactive, the tracing is disabled. 
% For example, we can configure a context ID comparator 
% to only generate a trace for a particular process, 
% because only when the context ID register
% holds that value can trace data be generated. 
Second, a code region can be included
or excluded from tracing. 
This is achieved by setting 
a pair of address comparators. 
Finally, it can be controlled 
by the \textit{trace start/stop block}. 
If an event happens, tracing is started. 
The tracing does not stop until the
block receives a stop signal, 
which is specified by another event resource.
Unfortunately, only the last method is supported by ARM MCUs~\cite{m4trm}.
Worse, the comparator resources, which are provided
by the DWT unit, are very limited.
This poses a significant challenge for us 
to effectively filter the execution trace we care about.

\section{\sys Design}
In this section, we begin with an overview of the \sys architecture,
and then delve into the detailed design of two critical components, 
the online trace collector and the offline trace analyzer, 
as well as their interaction with the AFL framework.

\subsection{Overview}

\begin{figure}[t]
\centerline{\includegraphics[width=\columnwidth]{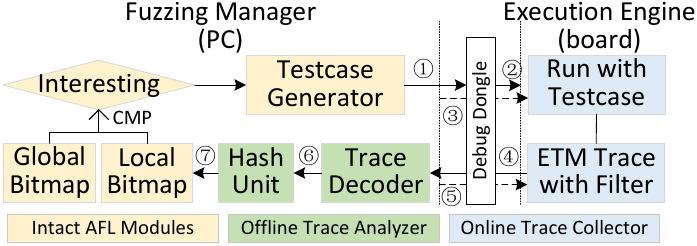}}
\caption{\sys Overview}
\label{fig:architecture}
\end{figure}

\sys is a new fuzzing tool designed for MCU firmware,
with a focus on peripheral driver code.
The fundamental idea is to 
inherit the sophisticated genetic algorithm of AFL 
while replacing its process-based execution engine with two critical components,
an online trace collector and an offline trace analyzer, as shown in Figure~\ref{fig:architecture}.
This design enables \sys to test all the code in firmware,
including peripheral drivers and closed source libraries.
% The following describes the workflow of \sys.
% Our work, standing on the shoulders of the mighty AFL,
% preserves the testcase generation component and 
% the bitmap analysis component to guide the whole fuzzing procedure.
% To bridge the on-device execution with
% the rest framework of AFL, we leverage a debug dongle,
% which is a must-have tool for embedded system development.
% Furthermore, \sys provides two innovative components
% \textit{online-trace-collector} and \textit{offline-trace-analyzer}
% to replace the execution engine part mentioned in Section~\ref{bg_afl}
% and assist the manager to collect the execution paths from a separate world
% and map them into the bitmap.

As illustrated in Figure~\ref{fig:architecture}, the host PC and target board communicate via a debug dongle,
a must-have tool for embedded system development.
To begin with, the host PC feeds the testcase 
via the debug dongle
into a reserved memory on the target board
(\textcircled{1}) and directs the target to begin execution (\textcircled{2}).
Once the target firmware has reached the point where testcase is firstly consumed,
the host PC sends the command
to activate the ETM function (\textcircled{3}).
Then, while the firmware is executing, the generated instruction trace is synchronously streamed 
to the host PC via the debug dongle (\textcircled{4}).
After completing one round of execution, the host PC sends another command
to deactivate ETM (\textcircled{5}).
The collected trace information is then used to reconstruct
the execution paths.
\sys adopts a novel scheme for representing branch edges (\textcircled{6}). 
The final result is mapped into the bitmap
to determine whether a new path has been discovered
and guide the generation of new testcases following the same genetic algorithm of AFL (\textcircled{7}).

% We will detail the design of two critical components in the following subsection: the online trace collector and the offline trace analyzer.

\subsection{Low-level Device Control and Fuzzing Scheduling}
\label{sec:coordination}
We use the JTAG or SWD interface for low-level control of the target device.
Through these interfaces, the debug dongle can
directly access the processor registers and the device memory (including
the memory-mapped system configuration registers) via the Debug Access Port~\cite{debugport}.
It also gives us the lowest control over the target device.
This is important because fuzzing often causes the target device
to enter a non-responsive state. If this happens, we can
force a reset via the low-level JTAG/SWD command without human involvement.

We also need to send testcases to the target device. Depending on the generation algorithm,
the size of a testcase can be as large as several megabytes. We leverage the
SEGGER  RTT (real time transfer) protocol~\cite{rtt} for high-speed transmission.
Under the hood, RTT uses AHB-AP (Advanced High-performance Bus - Access Port)~\cite{ahb}
to access memory in the background.
Not only can it provide enough bandwidth,
but also enables parallel scheduling.
Specifically, we transmit the subsequent testcase in the background
while the target is running against the current testcase.
Moreover, we conduct the analysis of the previous testcase on PC
in parallel with the target execution.
By scheduling all the tasks in a pipeline, \sys achieves optimized performance.

% First, performing fuzzing between two distinct worlds requires 
% the host PC to communicate with the target board
% to manage the fuzzing stages.
% The host PC, however, has no direct method of doing so.
% Fortunately, 
% ARM SoCs provide a number of CoreSight components for developers 
% to monitor, debug, and optimize the firmware. \sys leverages these components and the external debug dongle to reset the device, activate/deactivate ETM, feed testcase and suspend/resume its execution. (\gl{This paragraph is not solely for the purpose of feeding test cases, and thus I believe it should be moved ahead.})

\subsection{Online Trace Collector}
The online trace collector is responsible for collecting the ETM 
instruction trace during the firmware execution. 
It needs to solve two major challenges: 
1) how to feed testcases to the target board from PC,
2) how to selectively collect a minimal but sufficient instruction trace of
interesting code snippets.
% Following that, we will discuss the novel designs of \sys that address both challenges.

% \begin{enumerate*}[label={\alph*)},itemjoin={{, }}, itemjoin*={{, and }}, font={\color{red!50!black}\bfseries}]
%     \item how to feed the various testcases from a separated world
%     \item how to generate the instruction trace efficiently.
% \end{enumerate*}
% 1) how to feed the various testcases from a separated world,
% and 2) how to generate the instruction trace efficiently.

\subsubsection{Testcase Feeding}
\label{sec:runwithtc}
We reserve two fixed arrays to hold testcases,
one for the current test and the other for the
subsequent test, which is transmitted in the background during
the current test to improve parallelism.
The communication channel being used is SEGGER RTT as mentioned before.
These arrays are declared in a \texttt{noinit} section so that
the libc constructors will not interfere with them during initialization.
The size of arrays can be configured as needed.

% When feeding the testcase to the firmware, it typically does not have any dedicated RAM space for it. 
% Therefore, we need to reserve a contiguous memory space to hold the testcase. 
% This could be accomplished by creating an array or reserving it in the link script during taming.
% Meanwhile, the host PC should be unable to write to the reserved memory until the firmware's libc completes the execution of \textit{ctors}, 
% a collection of execution startup routines linked into a C program, 
% to avoid data collisions caused by concurrently writing to the reserved testcase buffer.
% As a result, \sys designates the reserved testcase buffer as a \textit{noinit} section,
% which is not initialized to zero at startup,
% preventing \textit{ctors} from resetting that region. 
% Finally, \sys makes use of the debug dongle to feed the testcase through
% the Advanced High-performance Bus (AHB)~\cite{ahb}, when the firmware is running
% from the \textit{ResetHandler} to the fuzzing start point to gain better performance.
% This enables parallel execution of the firmware and writing of the testcase to the reserved memory.

\subsubsection{Trace Collection and Filtering}
\label{sec:etmtracefilter}
To collect the execution trace, we can instrument the firmware so that
each basic block transition can be recorded and streamed to the PC. However, two
challenges need to be addressed.
% \begin{enumerate*}[label={\alph*)},itemjoin={{, }}, itemjoin*={{, and }}, font={\color{red!50!black}\bfseries}]
\begin{enumerate*}[label={\alph*)},itemjoin={{, }}, itemjoin*={{, and }}]
    \item The instrumentation requires additional memory space and computing
    resources that resource-restricted MCU chips may not afford
    \item The current binary rewriting techniques still face some fundamental technical 
    issues (\eg~current disassemblers cannot disambiguate between references and literal values 
    precisely), especially when the target binary is stripped.
\end{enumerate*}

% AFL instruments random numbers at each basic block of the program
% and writes them into a shared bitmap between AFL and the target program
% to indicate different paths. Unfortunately, this scheme is incompatible with embedded systems for three reasons: 
% \begin{enumerate*}[label={\alph*)},itemjoin={{, }}, itemjoin*={{, and }}, font={\color{red!50!black}\bfseries}]
%     \item the instrumentation requires a large flash space that resource-restricted embedded systems may not afford
%     \item the host PC cannot share memory with the embedded system
%     \item such instrumentation introduces additional performance overhead. 
% \end{enumerate*}

To tackle this problem, \sys leverages the ETM hardware feature to generate the instruction trace.
The collection is transparent to the firmware.
Therefore, no software instrumentation is needed and no additional overhead
is incurred.
% When the firmware is running, ETM generates trace packets automatically. 
By default, ETM collects all the branch information which
is sufficient to recover the full instruction trace of a testcase.
However, based on our experiments, this is sub-optimal (see \textbf{RQ2} in Section~\ref{sec:evaluation}).
In particular, lots of irrelevant packets have to be
transmitted and analyzed.
For example, the booting process of an MCU is fixed and never influenced
by a testcase.
We can safely avoid collecting ETM data during device booting to save resources.
Moreover, even if we have the abundant resources to do so,
the irrelevant packets add noises to the fuzzer that cannot be easily removed.
For example, some MCU firmware is multi-tasked. Collecting
all the trace information means all the tasks are traced.
This brings about non-determinism that leads to a different trace at each run
even if the testcase is the same.

% is imposes a significant overhead, 
% as transmitting and decoding a large number of packets requires significant time 
% on both the debug dongle and the host PC. 
% \begin{enumerate*}[label={\alph*)},itemjoin={{, }}, itemjoin*={{, and }}, font={\color{red!50!black}\bfseries}]
%     \item more packets will incur longer time to transfer them to the host PC
%     \item more packets will introduce more decoding operation on the host PC
%     \item the meanless and tedious firmware code snippets are helpless for fuzzing
%     \item even if only part of packets is enough to identify different paths.
% \end{enumerate*}
% Given that only a subset of packets is required to distinguish between different paths, 
% we use the DWT hardware feature to support filtering unwanted ETM data.
% \sys designs two kinds of online filters by leveraging
% DWT.
% The filters can reserve those trace packets of key code snippets
% and filter out useless packets while introduces negligible
% performance overhead and reduces the noise of the bitmap.
We use the DWT hardware feature to filter out irrelevant ETM packets.
However, in ARM MCUs, DWT only implements a limited number of
comparators (four in a typical implementation) that can be used as filters.
We have to prioritize its usage to maximally reduce irrelevant packets.
Based on the ETM triggering methods mentioned in Section~\ref{sec:dwt},
% the filtering conditions
% in order to minimize unwanted ETM traces
% and focus the collection on the relevant code areas.
% \sys cannot arbitrarily filter any 
% \gl{\sys cannot achieve infinite trace filters}, 
% we still designed the trace filters with our 
% best effort to make \sys can collect trace packets of key code
% as many as possible and hence to discover more paths.
we design two kinds of online filters, namely \textit{address-based filter}
and \textit{event-based filter}. 
When an MCU has more comparator resources, these filters
can be combined to generate more fine-grained traces.
% Besides, if additional DWT comparators are implemented 
% on the ARM Cortex-M in the future, 
% \sys can be easily extended to provide more configurable sets of both filters
% and generate finer-grained filtered instruction trace.

\paragraph{Address-based Filter.}
This filter allows analysts to specify a continuous code region
to be traced. It works the best when we are interested in a particular library.
This mode consumes two comparators to configure the region start and end.

\paragraph{Event-based Filter.}
\sys also supports event-based filters in which certain events trigger
the on/off switch of ETM. The event can be either executing an instruction
in a particular memory range or reading/writing a particular value from/to a particular address.
Both consume two comparators.
We call the former instruction trigger and the latter data trigger.

The instruction trigger is very useful in skipping the device booting process.
We use the code snippet in  Listing~\ref{lst:instaddronly} as an example.
Lines 4-7 are part of device booting and they have nothing to do with the testcase.
Line 12 is the main logic of the firmware, which is put in an infinite loop.
This is the paradigm in MCU programming -- the main operation is executed constantly
to sense environmental data and process them accordingly in a loop.
We tame the code by adding three lines (9, 13, 14).
Note although \sys does not require the source code of the target library,
the source code that invokes the target library is needed to
make it easy to tame the fuzzing process.
\texttt{fuzz\_stop} is a flag that marks whether the fuzzing should stop.
It can be changed by the firmware when the testcase has been used up or
by the debug dongle asynchronously.
By configuring the instruction address at line 9 to start ETM
and configuring the instruction address at line 14 to stop ETM,
\sys can effectively focus on the main logic of the firmware.

\begin{lstlisting}[language=c, label={lst:instaddronly}, caption={A code snippet containing
initialization code, main application logic, and \sys harness}]
int fuzz_stop = 0;
int main(void)
{
    MPU_Config();
    SCB_EnableICache();
    SCB_EnableDCache();
    HAL_Init();
    ...
    fuzz_stop = 0;
    while (1)
    {
        MX_USB_HOST_Process();
        if(fuzz_stop)
            break;
    }
}
\end{lstlisting}

% we do not need to trace the execution of the initialization code, 
% such as the code of configuring the MPU, cache, 
% and HAL (\ie~hardware abstract layer),
% because it is irrelevant to the testcases and
% their execution always generates the same
% path, which may further cause conflicts with other useful paths when
% generating the bitmap.
% \begin{enumerate*}[label={\alph*)},itemjoin={{, }}, itemjoin*={{, and }}, font={\color{red!50!black}\bfseries}]
%     \item the execution of them is independent of the testcases
%     \item the execution of them always generate the same path
%     \item these paths may cause conflict with other useful paths when generating the bitmap.
% \end{enumerate*}
% Therefore, \sys specifies a start point and an end point
% (represented as the instruction addresses)
% to trace a part of the firmware.
% % trace start point and trace end point through the instruction address.
% As shown in Listing~\ref{lst:instaddronly},
% \sys can configure the trace as starting at line 9 and ending at line 14 to
% trace the USB host driver related code. 

The data trigger provides fine-grained tracing capability, which we leverage
to trace a specified task.
More specifically, in the multi-task environment,
we use the  data trigger mode to 
filter out the execution trace of other tasks and the OS kernel, such as
interrupt handlers and scheduling, which is considered as noise to the fuzzer.
We observe that in the RTOS environment, each task has its
task control block (\textit{TCB}) in a fixed location regardless
of the testcase being used.
Therefore we can configure DWT to turn on ETM
when the global pointer 
that points to the TCB of the current task (\eg~\texttt{pxCurrentTCB} in FreeRTOS)
is written with the TCB address of the target task.
% All the schedulers switch tasks through this method.
When any other value is written to that pointer, indicating
the target task is swapped out, ETM is turned off.

\subsection{Offline Trace Analyzer} 
\label{sec:offline}

For each testcase execution, the offline trace analyzer processes the ETM
packets from the target device. It first recovers the branch
information without decoding the raw ETM data. This is achieved using a kind of special basic block.
Then, the branch information derived from
the basic block transitions is used 
to update the bitmap of code coverage maintained by AFL. 
All the other AFL components are intact, including
the new path identification module, the genetic algorithm to generate new testcases, etc.

%  which leverages the bitmap to identify interesting
% testcases and use genetic algorithms to generate promising testcases.

% Once a round of execution is completed,
% the offline trace analyzer is responsible for identifying the execution paths
% and mapping them to the bitmap, which is then used to guide the next fuzzing round through \textit{Trace Decoder}
% and \textit{Hash Unit}, respectively. 
% \sys provides a \textit{Trace Decoder} and a \textit{Hash Unit}
% to map the raw instruction trace to the bitmap. 
% To reflect the bitmap with the raw ETM packets, 
% \sys offers one Trace Decoder unit and one \sys Hash unit, as described below.

\begin{figure}[t]
\centerline{\includegraphics[width=0.7\columnwidth]{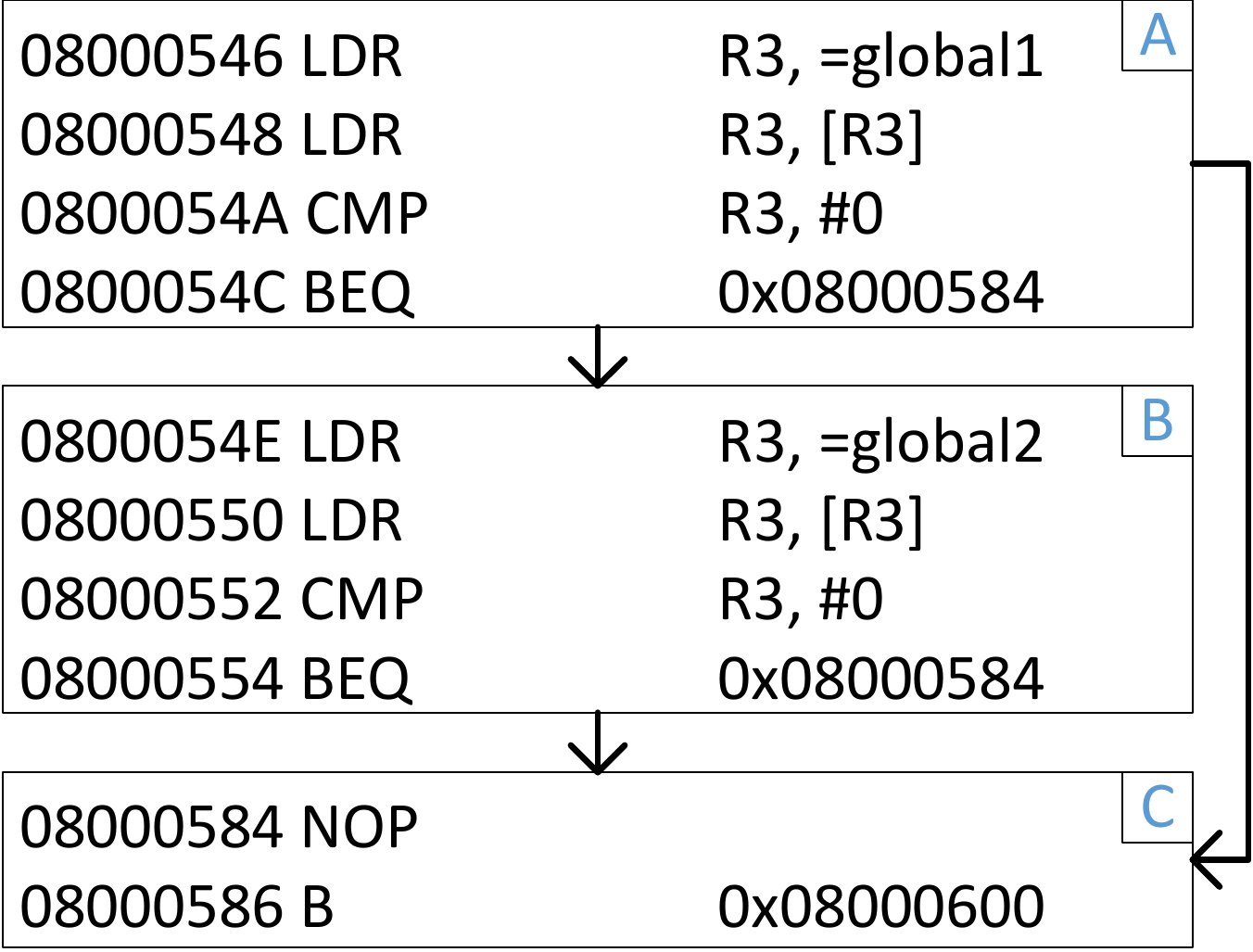}}
\caption{Control flow graph of an example code}
\label{fig:dbb_code}
\end{figure}

\subsubsection{Calculating Branch Information without Decoding ETM}
\label{sec:traceanalysser}
% When the filtered trace is transferred to the host PC,
% \sys will 
AFL uses the branch (edge) coverage information to identify
novel execution paths.
% whose testcase is marked as interesting for the mutator to generate new testcases.
To capture branch information, AFL (in QEMU mode) dynamically captures
basic block transitions and uses the address of basic blocks to populate
the bitmap following the code below~\cite{aflwhitepaper}.

\begin{lstlisting}[language=c, label={lst:aflqemu}, caption={Branch coverage calculation in AFL QEMU mode.}]
cur_location=(block_address>>4)^(block_address<<8);
shared_mem[cur_location^prev_location]++; 
prev_location=cur_location>>1;
\end{lstlisting}

In Listing~\ref{lst:aflqemu}, \texttt{shared\_mem} refers to the local bitmap of the current testcase
and \texttt{block\_address} is the address of the current basic block.
In line 1, \texttt{block\_address} is fed to a simple hash function
to get a random identification of the current basic block, denoted as \texttt{cur\_location}.
For the path $A \rightarrow B \rightarrow C$ shown in Figure~\ref{fig:dbb_code},
the basic block transition sequence is
$0x8000546 \rightarrow 0x800054E \rightarrow 0x8000584$,
and for the path $A \rightarrow C$,
the basic block transition sequence is $0x8000546 \rightarrow 0x8000584$.
Following the code shown above,
these two paths generate two different bitmaps,
which AFL can leverage in finding interesting testcases.

Using ETM trace, we can recover the same branch coverage information.
Concretely, by walking along the disassembled instruction sequence and aligning
it with decoded ETM packets,
we can recover the whole instruction trace and further rebuild the same branch
information as AFL.
However, this incurs non-trivial overhead according to the literature~\cite{ge2017griffin}
and it is also verified by our experiments (see Section~\ref{sec:decodeETMperformance}). 

To avoid expensive code disassembling and ETM decoding,
we propose a novel mechanism to capture branch coverage information directly using
the raw ETM packets.
A straightforward idea is to use the target address of every ETM branch packet
as the start of each basic block.
However, it cannot differentiate the two paths in Figure~\ref{fig:dbb_code},
because the conditional branches at addresses $0x800054C$
and $08000554$ do not
generate any branch packet no matter the branch is taken or not.
% (recall that P-header is used to record taken-or-not-taken information).
Fortunately, we found in the ETM manual~\cite{m4trm} that 
this behavior can be overridden by setting the eighth bit of the \textit{ETMCR} register.
In particular, with this option enabled, ETM will 
generate branch packets for direct branches that are actually taken.
% by setting the eighth bit of the \textit{ETMCR} register.
% A caveat is that the branch packet is only generated when the branch is actually taken.
Therefore, the
path $A \rightarrow B \rightarrow C$ would emit a sequence of ETM packets
($0x8000546, EEENEEEE, 0x8000584$),
while the path $A \rightarrow C$ would emit a sequence of ETM packets
($0x8000546, EEEE, 0x8000584$).
Here, $E$ is a bit in the P-header which means the condition of an instruction is true while
$N$ means the opposite, as mentioned in Section~\ref{sec:bg:etm}.
As can be seen, the latter path directly jumps from the basic block $A$ to $C$
since the branch condition at address $0x800054C$ is true.
A branch packet with target $0x8000584$ is thus emitted following the $E$ bit.
On the contrary, for the former, the branch condition at address $0x800054C$ is false
and therefore no branch packet is generated there.
However, the branch condition at address $0x8000554$ is true,
which leads to a  branch packet with target $0x8000584$.
By comparing the two ETM traces,
it is obvious that we can differentiate the two paths
since the P-header bits in between the two branch targets are different.

To explain the branch coverage information \sys captures,
we first explain a kind of special basic block generated
with linear code sequence and jump (LCSAJ) analysis~\cite{yates1995effort}.
We call it \BB.
A \BB is \emph{an instruction sequence starting from the last taken branch target
and ending with the following branch instruction which is actually taken}.
Under this definition, the basic block \textit{A} alone
is a \BB in the path $A \rightarrow C$,
while in path $A \rightarrow B \rightarrow C$,
the basic block \textit{A} concatenated with basic block \textit{B} constitutes
a \BB since the branch at the end of $A$ is not taken.
In \sys, we represent a \BB by combining the base address obtained from
the previous branch target and the P-header bitstream before
the next \BB, formally donated as $(BB\_base, BB\_bitstream)$. For example,
the \BB for $A$ in path $A \rightarrow C$ is encoded as ($0x8000546$, 1111),
while the \BB for $A|B$ in path $A \rightarrow B \rightarrow C$ is encoded as ($0x8000546$, 11101111).
Note this information can be obtained without referring to the assembly code.
In Section~\ref{sec:hash}, we explain how to use \BB 
transitions to calculate branch coverage to bridge
with AFL.
% so that the bitmap can be maintained.
It is worth noting that our approach does not generate the same
bitmap as AFL does. However, it achieves the same path sensitivity
since any change in basic block transitions will be reflected on the change 
in the corresponding \BB (either $BB\_base$ or $BB\_bitstream$).

\paragraph{Nondeterminism.} Although the online trace collector can already
 filter out a substantial amount of relevant ETM packets, limited by the
 hardware, it still emits many noisy packets. For example, there is no
 mechanism to suppress the tracing of exception handlers, which happen
 non-deterministically. This will add instability to the fuzzer because the
 same testcase would generate different execution traces in different runs. To address the
 issue, \sys also provides an offline exception filter. Specifically, we
 leverage the exception information embedded in the ETM branch packets to
 figure out the exception entry points and exit points. The analyst can
 choose whether or not to discard the ETM trace generated during the handler execution. Again,
 no disassembling is needed in this process.

\subsubsection{Mapping Branch Information to the Bitmap}
\label{sec:hash}

In this section, we explain how to map \BB-based branch information to the
bitmap maintained by AFL. 
Following the AFL design shown in Listing~\ref{lst:aflqemu},
our goal is to transform a \BB denoted 
by $(BB\_base, BB\_bitstream)$ into a unique number,
which will be used to replace of role of \texttt{cur\_location}
(unique identification for the basic block)
in Listing~\ref{lst:aflqemu}.
To sufficiently diffuse the information contained in each \BB,
% Initially, we na\"ively used an algorithm similar
% to the one listed in Listing~\ref{lst:aflqemu}. However, we observed
% significant collisions in the bitmap. The reason is that the address space
% for the code area in MCUs is small, leading to a significant overlap in the
% most significant bits among different \textit{dynBBs}. Using the simple
% shifting operation in line 1 of Listing~\ref{lst:aflqemu} cannot sufficiently
% diffuse the address and thus leads to collisions in the bitmap. We address this problem by designing 
we adopt a lightweight hash algorithm based on
MurMurHash~\cite{murmurhash}.
The output is a random integer $BB\_ID$.
As mentioned before, the whole algorithm replaces line 1 of Listing~\ref{lst:aflqemu}.
We split the bitstream into chunks of 5-bits and apply bit-wise
XOR on them, yielding a number $t$ ranging from 0 to 31 (\ie~\texttt{FoldAndXor()}). 
Then, $t$ is used to mix with and shift $BB\_base$.
The result is further split into two parts and mixed with some magic numbers.
% It replaces the len variable in Algorithm.1.
While this design is ad-hoc,
it effectively randomizes the encoded \BBs such that 
the resulting $BB\_ID$ does not incur too many
collisions on the AFL bitmap based on our evaluation.

\begin{algorithm}
\SetAlgoLined
\KwIn{\textit{(base,~bitstream)} $\gets$ ($BB\_base,~$BB\_bitstream$)$ \newline %0x0~0x100000
\textit{MAP\_SIZE} $\gets$ length of bitmap in bytes} 
\KwOut{\textit{BB\_ID}}
 \SetKwFunction{FMain}{HASH}
 \SetKwProg{Fn}{Function}{:}{}
    \Fn{\FMain{$base, bitstream$}}{
        $t \gets FoldAndXOR(bitstream)$; \\ 
        $base \gets base + t$; \\ 
        $left \gets (base \ll (32-t)) \mid (base \gg t) $; \\
        $right \gets (base \ll t) \mid (base \gg (32 - t))$; \\
        $BB\_ID \gets (left \mid right)$; \\
        $BB\_ID \gets (BB\_ID \oplus (BB\_ID \gg 16))$;\\
        $BB\_ID \gets (BB\_ID * 0x85ebca6b)$;\\
        $BB\_ID \gets (BB\_ID \oplus (BB\_ID \gg 13))$;\\
        $BB\_ID \gets (BB\_ID * 0xc2b2ae35)$;\\
        $BB\_ID \gets (BB\_ID \oplus (BB\_ID \gg 16))$;\\
        $BB\_ID \gets ((BB\_ID \gg 4) \oplus (BB\_ID \ll 8))$;\\
        $BB\_ID \gets (BB\_ID \wedge (MAP\_SIZE - 1))$;\\
        \textbf{return} $BB\_ID$;
    }
    \textbf{End Function}
 \caption{Hash function to transform a \BB into a random ID}\label{alg:hash}
\end{algorithm}

\subsection{Crash/Hang Detection}
\label{sec:abnormalstatuscapture}

% Due to lack of memory protection mechanisms, crash detection is challenging in
% embedded systems~\cite{corruptcrash}. 

\sys relies on the built-in exception handling mechanism to detect abnormal
firmware behaviors. Specifically, we use the vector catching feature~\cite{fsrfar} 
to mark the exceptions in concerns, such as \textit{Hard Fault}, \textit{Mem Manage},
\textit{Bus Fault}, \textit{Usage Fault}, etc. These exceptions indicate
critical system errors and thus can be used as crash signals. With vector
catching, when such an exception happens, instead of trapping to the
corresponding handlers, the chip  enters debug state which can be
automatically captured by the debug dongle. Then, \sys further checks the
Fault Status Registers and Fault Address Registers to examine the root
cause. To notify the fuzzing manager of a successful execution,
at the end of the tested code, we place
 a \texttt{BKPT} instruction with a magic number as the argument. If the current
 test terminates correctly, the execution will enter debug state at this 
\texttt{BKPT} instruction. We further check the value of the argument to confirm a
 successful execution. Lastly, if the fuzzing manager does not capture any
debug state in a specified amount of time (we use two seconds as an empirical value),
a hang is marked.

\begin{figure}[t]
\centerline{\includegraphics[width=.8\columnwidth]{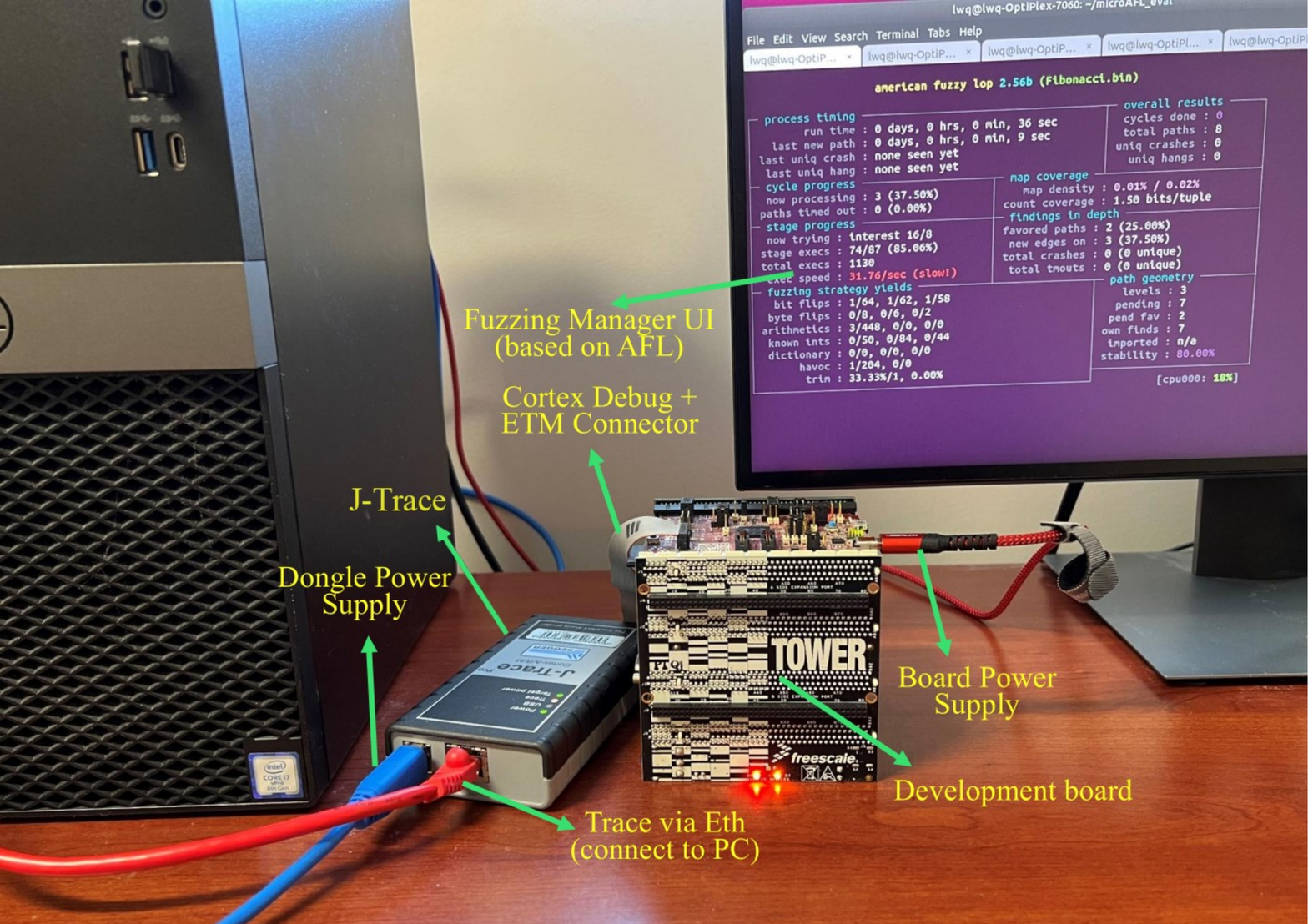}}
\caption{\sys setup in fuzzing a firmware sample on the NXP TWR-K64F120M board}
\label{fig:microAFL_setup}
\end{figure}

\section{Implementation and Evaluation}
\label{sec:evaluation}
% We evaluated the efficiency and contribution
% to the overall performance of each component.
% In addition, we evaluated the effectiveness of \sys upon
% the real-world peripheral vulnerabilities discovery capability.

We have implemented a prototype of \sys on top of AFL2.56~\cite{afl}
by adding $\sim$2,000 lines of C code on the PC side. 
% The \textit{online-trace-collector} relies on 
% ARM Cortex-M ETM and DWT hardware features
% to collect and filter instruction trace.
% The transmission of the control command and the data
% between the host PC and the target ARM Cortex-M board
% is based on the debug dongle SEGGER J-Trace Pro
% and the counterpart J-Link SDK.
We used the SEGGER J-Trace Pro debug dongle~\cite{seggerjtrace} to control the
communication between the host PC and the target ARM Cortex-M evaluation boards. 
The control logic was implemented on the PC side
using the SDK provided by SEGGER~\cite{seggersdk}.
In Figure~\ref{fig:microAFL_setup},
we show the setup in which we used our prototype to
fuzz a firmware sample for the NXP TWR-K64F120M evaluation board.
The key components are annotated corresponding
to the architecture diagram in Figure~\ref{fig:architecture}.

We evaluated the performance of our \sys prototype to answer five research questions.
The answers are expected to demonstrate the unique strengths of \sys and its effectiveness as a new fuzzing solution for embedded firmware testing.
%Evaluations are conducted to address the following research questions in order to substantiate our earlier claims and show a comprehensive result:
\begin{enumerate}
    \item[\textbf{RQ1:}] {\em Does the \BB-based approach that directly processes raw ETM trace improve performance compared to approaches that fully recover the instruction flow?} %{\em What is the performance of \sys when disassembly is used?} 
    \item[\textbf{RQ2:}] {\em Do online filters help to improve the performance of \sys?}
    % How much contribution do they provide?} 
    %How do online filters contribute to performance improvement?
    \item[\textbf{RQ3:}] {\em How much overhead does \sys introduce to (each sub-process of) a single round of fuzzing?} %How much overhead is involved in each step of a single round fuzzing?
    \item[\textbf{RQ4:}] {\em What is the overall performance of \sys and how is it compared with existing work?}
    \item[\textbf{RQ5:}] {\em How effective is \sys in locating bugs in the peripheral drivers of real-world firmware?}
\end{enumerate}

% To answer our research questions, we conducted the following experiments:
% \begin{enumerate}
%     \item[1)] Identify the basic blocks used by AFL with disassembly when rebuilding paths and compare performance to \sys.
%     \item[2)] Compare the performance of \sys with and without filters. 
%     \item[3)] Statisticize the execution time for each step in a single round.
%     \item[4)] Run \sys for an extended period of time to test the various firmwares.
%     \item[5)] Use \sys to fuzz different SDKs in order to uncover the real-world bugs.
% \end{enumerate}

% We evaluated \sys from two aspects.
% First, \gl{how many testcases can be executed per second?}
% Second, can \sys find bugs in real-world firmware,
% in particular, in peripheral drivers?
% the performance \gl{or efficiency??} and effectiveness of  by fuzzing
% the real-world firmware to find vulnerabilities in the
% peripheral drivers. 

\paragraph{Experiment Settings.}
The PC running the fuzzing manager is equipped with an Intel Core i7-8700 CPU@3.2GHz and 8 GB DDR4 RAM and SSD storage. 
We used the NXP TWR-K64F120M evaluation board as the execution engine and its corresponding SDKs for the performance experiments from \textbf{RQ1} to \textbf{RQ4}. 
To answer \textbf{RQ5}, we present a detailed case study about the real-world bugs that we found after a long-term fuzzing on the evaluation boards NXP TWR-K64F120M and STM32H7B3I-EVAL and towards various projects in their corresponding SDKs.
Both boards have ETM pinouts available.
% The debug dongle used was a SEGGER J-Trace Pro~\cite{seggerjtrace} which supports ETM tracing.
% calculates Fibonacci(1,000,000)
% without consuming any input

\paragraph{Firmware Samples Used in Evaluation.} For \textbf{RQ1} to \textbf{RQ4},
 we used the sample code provided in the NXP SDK.
 First, the sample \texttt{Fibonacci} implements a recursive function
 that calculates Fibonacci(1,000,000). This sample does not involve any peripheral
 and serves as the baseline in our evaluation.
 Second, the sample \texttt{I2C} involves the usage of a simple peripheral I2C.
 It merely communicates with the PC over the I2C bus.
 Third, the sample \texttt{UART} involves the usages of UART, which also
 communicates with the PC.
 Forth, the sample \texttt{USB} uses the evaluation board as the USB host
 to access a USB disk formatted as the FAT file system.
 Fifth, the sample \texttt{SD Card} recognizes and initializes a micro SD card inserted into the on-board
 SD card slot.
 Sixth, the sample \texttt{Enet} uses Ethernet to communicate with the PC over IPv4.
 Finally, the sample \texttt{MMCAU} uses the hardware Crypto Acceleration Unit (CAU)~\cite{mmcau}
 to complete cryptography operations such as AES, DES3, SHA, etc. 
 Note the library of CAU is closed source.
 We list the firmware information including the size and the number of basic blocks
 in Listing~\ref{lst:aflqemu}.
 For \textbf{RQ5}, we used the sample code provided in the NXP SDK and STM32 SDK.
 While we tested multiple drivers in both SDKs, we specifically
 chose USB as a case study to demonstrate the capability of \sys in finding
 bugs in complex peripherals.
 The application-level logic of all the samples is very simple, because
 we do not attempt to find bugs at high-level code.
 Rather, we make sure that the core peripheral functions
 are included in the sample, with the goal to feed abnormal inputs
 to the low-level driver code to trigger bugs.

\begin{table}
\adjustbox{max width=\columnwidth}{
  \begin{threeparttable}[t]
   \caption{Firmware size info and fuzzing performance under different settings (executions per hour)}
  \centering
   \begin{tabular}{l|r|r|r|r|r}
     \textbf{Sample} & \textbf{BB\#} & \textbf{Size (bytes)} & \textbf{Dis\&Dec}\tnote{*} & \textbf{\sys w/o Filter} & \textbf{\sys} \\
     \midrule
    Fibonacci& 9,927 & 12,064 & 100,830 & 33,386   & 104,394\\
    I2C      & 13,343 & 14,888 & 112,666  & 27,732  & 116,487\\
    UART     & 9,899 & 12,856 & 37,003  & 21,087    & 50,571 \\
    USB      & 36,431 & 40,024 & 465  & 2,211     & 2,236 \\
    SD Card & 21,104 & 25,328 & 592 & 2,207 & 2,335 \\
    Enet     & 14,475 & 18,504 & 685  & 1,089    	& 1,116 \\
    MMCAU    & 12,186 & 17,712 & 58,155  & 26,393    & 74,830 \\
     \end{tabular}
     \begin{tablenotes}
        % \item [1] Bytes are used to measure the size. 
        \item [*] Fully disassemble firmware
        and decode ETM data. The same filtering mechanism was applied as \sys.
     \end{tablenotes}
     \label{tab:ques_1_2}
  \end{threeparttable}
}
\end{table}

% \begin{table}[t]
%   \centering
%   \caption{Executions per hour under different settings.}
%     \begin{tabular}{l|r|r|r|r|r}
%     \textbf{Sample} & \textbf{BB num} & \tabincell{c}{\textbf{Size} \\ \textbf{(bytes)}} & \textbf{Disassy}\footnote{This solution fully recover the instruction flow with filter.} & \textbf{No Filter} & \textbf{\sys} \\
%     \midrule
%     Fibonacci& 9,927 & 12,064 & 100,853 & 33,386    & 104,394\\
%     I2C      & 13,343 & 14,888 & 112,666  & 27,732    & 112,931\\
%     UART     & 9,899 & 12,856 & 37,003  & 21,087    & 50,632\\
%     USB      & 36,431 & 40,024 & 465   & 2,211     & 2,236 \\
%     Enet     & 14,475 & 18,504 & 685    	& 1,089    	& 1,116 \\
%     MMCAU    & 12,186 & 17,712 & 58,155  & 26,393    & 74,832 \\
%     \end{tabular}%
%   \label{tab:ques_1_2}%
% \end{table}%

\subsection{Overhead of Fully Recovering Instruction Flow}
\label{sec:decodeETMperformance}

As discussed in Section~\ref{sec:traceanalysser}, \sys reduces processing overhead by
avoiding disassembling the firmware and aligning the ETM trace to recover
the full instruction flow. In this section, we demonstrate
how using the \BB-based approach can reduce performance overhead on PC.
%  identified the basic
% blocks used by AFL with disassembly when rebuilding paths and built a
% disassembly-supported \sys implementation. Then, we compared its performance
% with our \sys prototype to answer \textbf{RQ1}. 
In the implementation of the base line approach (fully
disassembling firmware and decoding ETM data),
we used the popular disassembly framework, Capstone~\cite{capstone}.
We followed the firmware execution by aligning the collected ETM packets with
disassembled instructions.
Whenever a new basic block was met, we disassembled the whole basic block at once,
which was also cached for future use.
After recovering the instruction trace, we followed
Listing~\ref{lst:aflqemu} to populate the bitmap.
It is worth mentioning that we applied the same filtering
strategy as \sys to ensure a fair comparison.
% Note that Capstone cannot distinguish between the instruction and data sections, due to a lack of symbol table information. So, it will terminate the disassembly process when encountering unknown bytes. Ideally identifying data mixed with the Thumb-2 instruction set in the firmware remains an unsolved challenge. As a result, Capstone cannot disassemble the entire firmware at once. Therefore, to ensure the paths rebuild process was successful, we had to align the instruction trace with the firmware and re-disassemble every time when it encountered a branch instruction.
We measured the total number of executions for each sample
in one hour. The results are shown in column 4 of Table~\ref{tab:ques_1_2}.
%  we compared the performance between
% disassembly-based \sys and \sys in seven projects. 
Compared with the performance of \sys (column 6),
we confirm that \sys is generally much
faster than the disassembly-based approach. 
% For firmware with more complex peripherals (e.g., \texttt{USB}), 
We observed 1.03x - 4.81x improvement on average.
%  From the
% observation, we conclude that disassembly imposes a significant overhead
% compared to \sys, which is path-sensitive as AFL while not requiring costly
% disassembly.

\subsection{Filter Performance}

To answer \textbf{RQ2}, we disabled the proposed online filters and compared
the performance of \sys with and without filters.
% For the same samples in Table~\ref{tab:ques_1_2}, 
As before, we recorded the number of executions within one hour.
The results are shown in columns 6 and 5 of Table~\ref{tab:ques_1_2} respectively.
First, we can see that online filters improve the performance of \sys in
general in all samples that we have tested. For very large samples
(e.g., the Enet, USB and SD Card), the
filter improves the performance of \sys slightly. However, for less
complex samples such as the MMCAU, the improvement
becomes very significant. This is because without filters, ETM needs to
collect the entire instruction trace beginning from \texttt{ResetISR} to the end of
each run, which not only increases the burden of the debug
dongle in transferring the raw trace, but also requires more time for the
offline decoder to decode and map them into the bitmap.
For larger samples, since the execution time is longer for each run,
the overhead can be amortized in the entire execution.

\begin{figure}[t]
    \centerline{\includegraphics[width=.8\columnwidth]{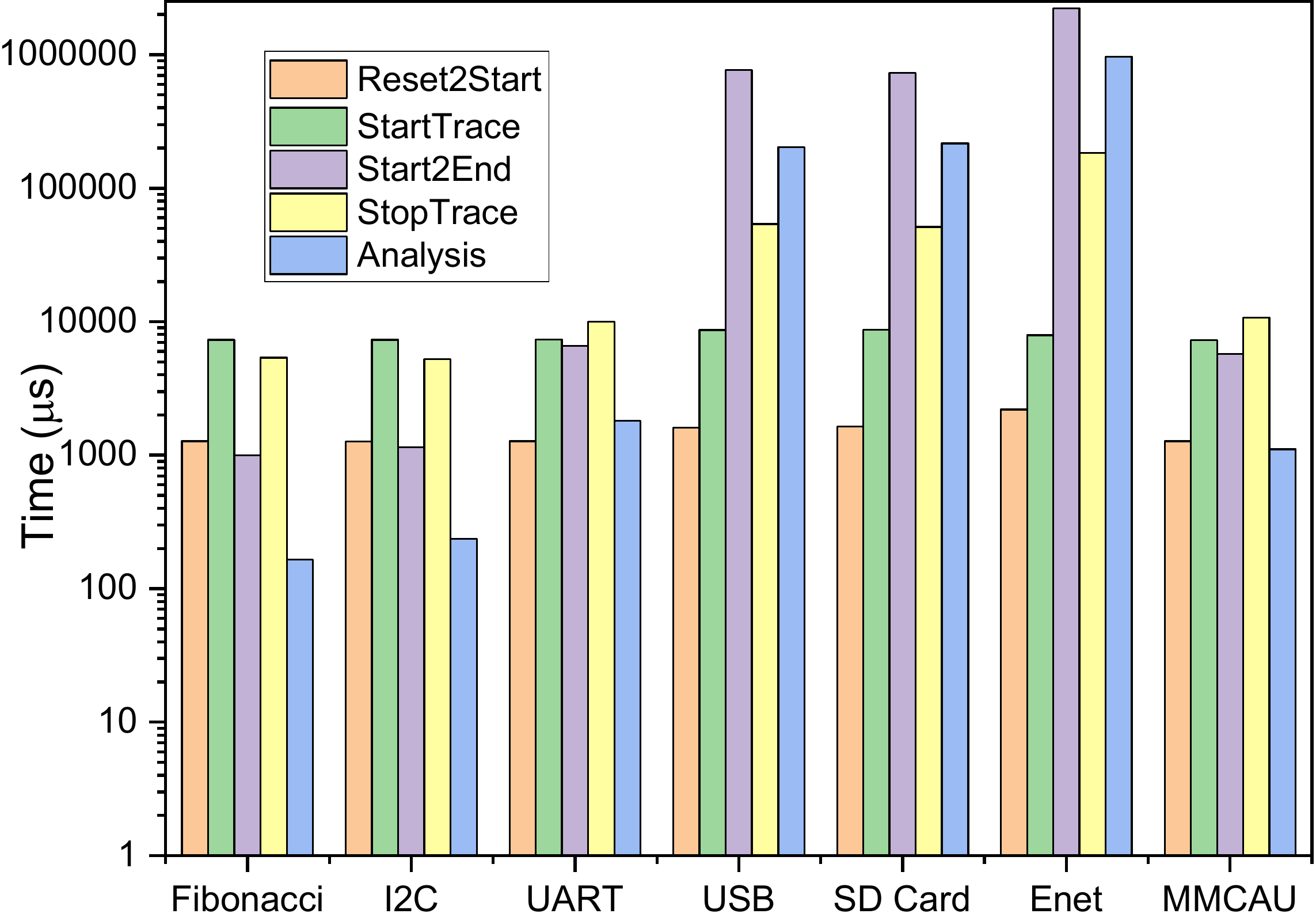}}
    \caption{Overhead breakdown of \sys (y axis is in logarithmic scale).}
    \label{fig:comp_perf}
\end{figure}

% \begin{figure}[t]
%     \centerline{\includegraphics[width=.8\columnwidth]{./img/component_proportion}}
%     \caption{Weight of each component for different firmware images.}
%     \label{fig:comp_perf_prop}
% \end{figure}

\subsection{Overhead Breakdown}

Each fuzzing round consists of five
sub-processes. \textbf{\em (1) Reset2Start} measures the time from device booting
(\ie~the start of \textit{Reset\_Handler}) to the fuzzing start point (\ie~start reading the testcase). 
Note we did not collect the branch coverage information during this sub-process
since it remains the same among all the testcases.
\textbf{\em(2) StartTrace} measures the time spent
on preparing a test, including the transmission of the testcase
from PC to the board and configuration of ETM and DWT functions.
\textbf{\em(3) Start2End} is the time spent from the fuzzing
start point to the end point. The branch coverage information is collected for this sub-process.
\textbf{\em(4) StopTrace} is the time spent on 
disabling the ETM trace and waiting for the completion of the ETM trace transmission.
\textbf{\em (5) Analysis} is time spent on analyzing the ETM packets.
% decodes the ETM packets and maps the paths into the bitmap.

% \begin{enumerate}
%     \item[1)] \textbf{Reset2Start.}
%     The firmware executes from
%     the \textit{Reset\_Handler} to the fuzzing start point.
%     \item[2)] \textbf{StartTrace.}
%     The host PC lets the SEGGER J-Trace Pro
%     prepare the transmission buffer and enables the ETM trace
%     functionality via the SEGGER SDKs API while the target board is suspended.
%     \item[3)] \textbf{Start2End.}
%     The firmware executes from the fuzzing 
%     start point to the end point. And the host PC collects the instruction 
%     trace transferred from the SEGGER J-Trace Pro synchronously.
%     \item[4)] \textbf{StopTrace.}
%     When the firmware finishes one round, the host PC deactivates the ETM trace 
%     functionality and awaits the completion of the rest trace transmission. 
%     \item[5)] \textbf{Analysis.}
%     The host PC decodes the ETM packets and
%     map the paths into the bitmap.
% \end{enumerate}

In this experiment, we measured the execution time of each sub-process to
assess its impact on the overall overhead. We calculated the average
execution time over 1,000 rounds of executions for each sample. 
Figure~\ref{fig:comp_perf} presents the breakdown of the execution time.
% whereas Figure~\ref{fig:comp_perf_prop} shows the overhead
% proportion of each subprocess. 
% as indicated in Figure~\ref{fig:comp_perf}
% to illustrate the effectiveness of each component.
% These projects were selected based on
% varied complexities and if is closed-source,
% to provide detailed information for future analyses
% and cover different real-world occasions.We divided the MCU firmware into four types as follows:
We can see that the execution time for the {\em Reset2Start} and
{\em StartTrace} sub-processes stay stable in all samples, while the
execution time for the other sub-processes increases as the firmware becomes
more complex. This is because more ETM packets were generated in more
sophisticated firmware samples. This will increase the overhead of not only
the \textit{StopTrace} operations due to waiting for the completion of the
trace transmission, but also the \textit{Analysis} operations due to the
increased decoding and hashing computation. We also observed that the execution
is very fast for simple firmware. For them,
the overhead is dominated by the 
\textit{StartTrace} and \textit{StopTrace} sub-processes, which are black-box
functions provided in SEGGER SDK. As the firmware becomes more complex, it takes
longer to complete the \textit{Start2End} sub-process and analyze the massive
instruction trace.

% Moreover, as shown in Figure~\ref{fig:comp_perf_prop}, the MCU execution is very fast for simple firmwares ({\em {\bf G1}}-{\em {\bf G3}}), therefore, the overhead is dominated by the ones introduced in the \textit{StartTrace} and \textit{StopTrace} subprocesses, which are black-box functions from the SEGGER SDK. As the firmwares become more complex, 
% it takes longer to complete the \textit{Start2End} subprocess and analyze the massive instruction trace.

% In summary, the breakdown of \sys is generally lower than 30\% and acceptable
% even when applied to the most complex peripherals.

\begin{table}[t]
  \centering
  \caption{Fuzzing performance}
  \resizebox{\columnwidth}{!}{
    \begin{tabular}{l|r|r|r|r|r}
    Project   & \multicolumn{1}{c|}{Time(s)} & \multicolumn{1}{c|}{Executions} & \multicolumn{1}{c|}{Exec/sec} & \multicolumn{1}{c|}{Paths} & \multicolumn{1}{c}{Crashes/Hangs} \\
    \midrule
    I2C  & 172,893 & 5,340,552 & 30.8893 & 2 & 0/0 \\
    UART  & 172,838 & 1,947,980 & 11.2706  & 27  & 1/0 \\
    USB   & 172,803 & 148,398   & 0.8588  & 201 & 0/500 \\
    SD Card & 171,693 & 149,063 & 0.8682 & 53 & 0/0 \\
    Enet  & 174,675 & 52,961    & 0.3032  & 84  & 14/0 \\
    MMCAU & 173,180 & 3,268,671 & 18.8744  & 95  & 0/0 \\
    \end{tabular}%
   }
  \label{tab:framework_efficiency}%
\end{table}%

\subsection{Overall Performance of \sys}
% !!!Please copy the table and edit it by sublime 
% or other text editors to get a better view

We evaluated the efficiency of the \sys framework in fuzzing
real-world firmware samples to answer {\bf RQ4}. We ran each firmware
for around two days and recorded the number of executions, the total
execution time, the number of covered paths, and the number of crashes and
hangs, as shown in Table~\ref{tab:framework_efficiency}.

% Overall, the results indicate that the performance of \sys is desirable for testing embedded firmwares, particularly with regard to its support to testing peripheral drivers. For example, it takes only ...
% \gl{add some explanation about the results}

We would also like to point out that, when the generated trace is very large
(e.g., over 200 MB), the debug dongle  became
insufficiently reliable. The buffer inside 
the debug dongle might be depleted and this led to
overflow and trace loss. If this happened, we had to force quit the execution
and reset the faulty cycle. This resulted in a lower
execution rate since we have to discard the faulty executions.
In rare cases, the error in the debug dongle became undetectable.
If that happened, we also observed some false positives.
This explains the false positives in the table.
When we replayed the same testcases that triggered crashes/hangs during fuzzing, 
the results could not be reproduced.

\begin{table}[t]
  \centering
  \caption{Comparison with related work (executions per hour)}
  \setlength{\tabcolsep}{3mm}{
    \begin{tabular}{l|r|r|r|r}
    \textbf{Sample} & \textbf{P$^{2}$IM} & \textbf{$\mu$Emu} & \textbf{Avatar$^2$} & \textbf{\sys} \\
    \midrule
    Console & 139,860 & 29,513 & 1,766 & 8,261 \\
    Fibonacci & 209,478 & 904,617 & 2,623 & 8,670 \\
    \end{tabular}%
    }
  \label{tab:comp_emu}%
\flushleft
\scriptsize{
There is a huge performance discrepancy between P$^{2}$IM 
and $\mu$Emu for the two samples. We attribute
it to the different strategies in selecting forking points.
}  
\end{table}%

\paragraph{Comparison with Existing Work.} As explained in 
Section~\ref{sec:intro}, \sys is the first work that can efficiently fuzz peripheral
drivers for MCU devices. The most related work includes 
Avatar~\cite{avatar} which emulates the firmware in QEMU but forwards peripheral
operations to the real development board, and pure emulation-based solutions
such as {P$^{2}$IM}~\cite{P2IM} and $\mu$Emu~\cite{uemu}.
The original Avatar only leverages the real hardware to pass
the firmware initialization phase and then uses symbolic execution to analyze
the code that never accesses peripherals. 
Later, it was substantially re-engineered for
multi-target orchestration purpose in Avatar$^2$~\cite{avatar2}.
However, fuzzing is not
supported by default on both. 
Emulation-based approaches provide great scalability but
cannot guarantee sufficient fidelity for fuzzing driver code.
For example, firmware with complex peripherals cannot be booted.

In this section, we ignore fuzzing effectiveness, but focus on the raw
fuzzing speed achieved by \sys, Avatar$^2$, {P$^{2}$IM} and $\mu$Emu. 
We selected Avatar$^2$ in our experiments for its
active development and more friendly API design.
Two
samples were used. Apart from the \texttt{Fibonacci} sample mentioned before,
we evaluated a \texttt{Console} firmware sample which was also used 
in {P$^{2}$IM}~\cite{P2IM} and $\mu$Emu~\cite{uemu}.
It provides a simple interactive shell via the UART peripheral.
% and the \texttt{Fibonacci} is a
% pure computation firmware which calculates Fibonacci(1,000,000)
% without consuming any input. 
It is worth noting that we tried to use
the firmware with more complex peripherals such as USB or Ethernet, but none of
the related work can support them. Indeed, Avatar$^2$ does not support DMA which is
indispensable for complex peripherals. {P$^{2}$IM} and $\mu$Emu failed to
emulate USB or Ethernet and thus cannot boot the firmware. We augmented the
original Avatar$^2$ framework with fuzzing capability,
and also improved the stability of the JLink (the name of SEGGER's debugging solution) target 
to manage the development board and to coordinate the fuzzing process\footnote{\url{https://github.com/avatartwo/avatar2/pull/96}}, similar to what we did in \sys. The
firmware starts execution in QEMU and forwards the I/O requests to the
development board when needed. For {P$^{2}$IM} and $\mu$Emu, we directly
built the systems using the source code indicated in the original papers.
However, we made some tweaks in fuzzing \texttt{Fibonacci}. Specifically,
{P$^{2}$IM} and $\mu$Emu mark the end of an execution only when all the
bytes in a testcase are used up. This led to constant time-out in fuzzing
\texttt{Fibonacci}. We addressed this issue by manually inserting into the
source code many invocations of \texttt{getchar()} to emulate consuming
testcases.

The results are summarized in Table~\ref{tab:comp_emu}. Emulation-based
solutions outperform hardware-in-the-loop solutions significantly because of
the higher computation power. More importantly, they avoid the time-consuming
synchronization between the PC and the board. \sys on the other hand outperforms
Avatar$^2$ since it suffers less from synchronization. The gap is narrowed for
the computation-intensive tasks (Fibonacci) because they do not frequently
access peripherals and thus avoid synchronization.

\subsection{Real-world Firmware Fuzzing}
\begin{figure}[t]
\centerline{\includegraphics[width=0.8\columnwidth]{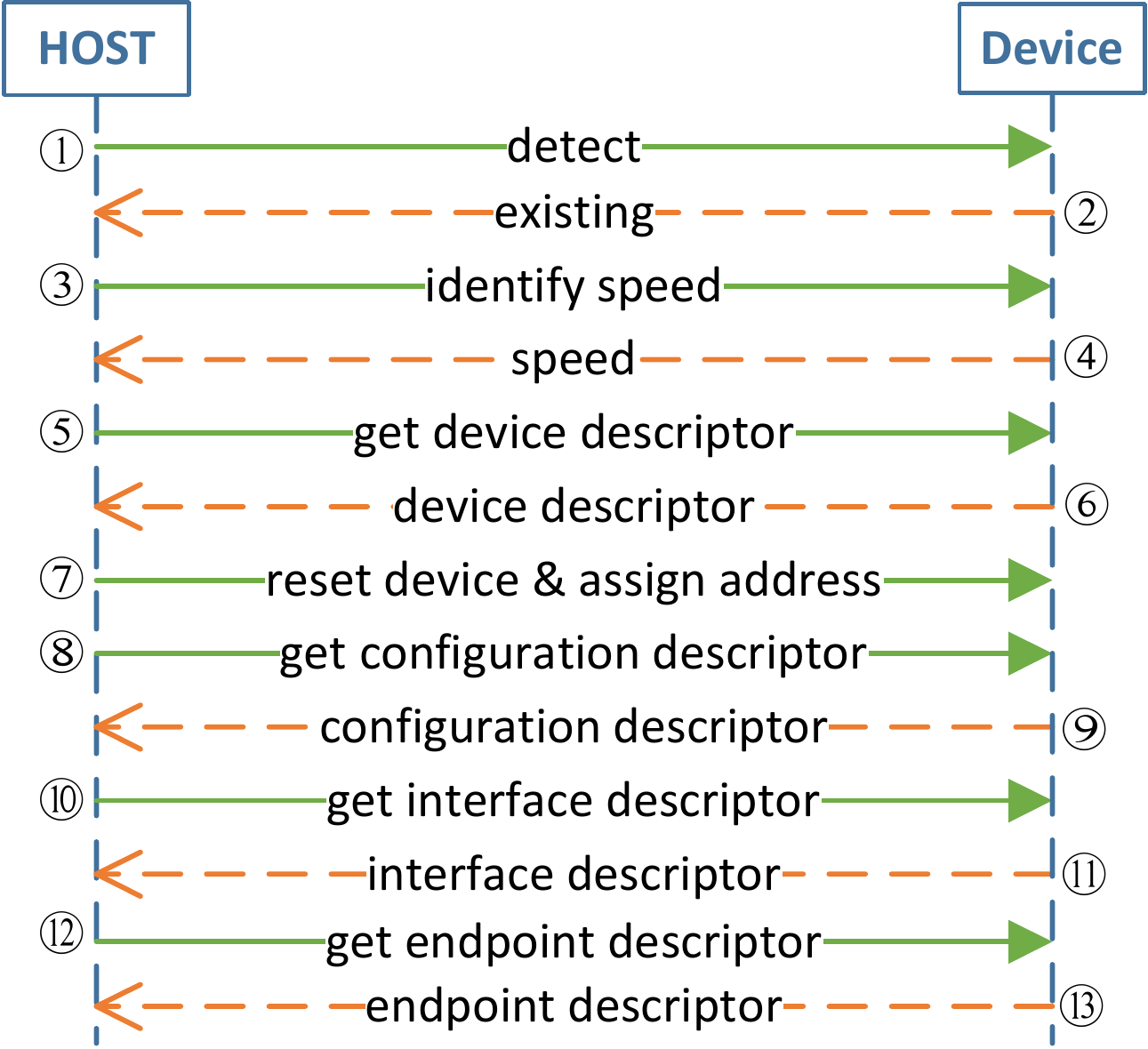}}
\caption{USB Enumeration}
\label{fig:usb_enum}
\end{figure}

To demonstrate the capability of \sys in finding bugs in the real world,
we fuzzed multiple samples with complex peripheral drivers for two days, including Ethernet, USB and SD card.
% \gl{We also tried to test the same samples using $\mu$Emu~\cite{uemu} and {P$^{2}$IM}~\cite{P2IM},
% two state-of-the-art MCU firmware fuzzing tools.
% However, both of them failed to run any of these samples.}
After running the fuzzer for two days for each sample,
we found ten previously unknown bugs in the USB drivers released with the STM32 SDK,
and three in NXP SDK.
In what follows, we use USB as a case study to show how \sys can efficiently
find bugs in complex driver code.

Our fuzzer focused on the USB enumeration process in which
the USB device is recognized by the host.
In our evaluation, we used the MCU board as a USB host and 
a SanDisk USB disk as a USB device.
We assume the USB disk is malicious and can send arbitrary data
to the MCU board.
We manually identified the program points where data are received
from the USB disk and replaced them with the testcases generated by AFL.

As shown in Figure~\ref{fig:usb_enum},
USB enumeration is very complex and involves multiple rounds of interactions.
Most importantly, the USB device is required to send
multiple descriptors to specify the device's properties.
% such as configuration descriptor, the interface descriptor and the endpoint descriptor.
When the USB host parses these descriptors, memory bugs could occur if
the relevant fields are not sanitized properly.
In our two-day fuzzing campaign,
we have
uncovered 13 previously undisclosed vulnerabilities.
All of them have been reported to vendors and patched.
Note that these bugs influence all the MCU devices
using STM32 SDK or NXP SDK.
% Based on the uncovered real-world bugs, we answer the \textbf{RQ5} that
% \sys is effective in discovering the vulnerabilities in the real world.
We categorize the bugs we found as follows.

% The USB protocol specifies the endpoint, 
% which is the hardware buffer in the device used for data transmission.
% On the host side, the endpoint is recognized as the software pipe.
% Vendors may manufacture USB devices with numerous optional endpoints
% but not less than one control endpoint, 
% known \textit{Endpoint0}.
% The host retrieves the device information or 
% sends the control commands through \textit{Endpoint0}.
% During enumeration, the host retrieves the device descriptor, 
% the configuration descriptor, the interface descriptor and the endpoint descriptor
% in succession as shown in steps \textcircled{5}, \textcircled{8},
% \textcircled{10}, \textcircled{12} of Figure~\ref{fig:usb_enum}.

% With intermittently fuzzing by \sys for two months,
% we have investigated different types of peripherals
% including USB, SD Card, Ethernet, MMCAU, UART, SPI, DAC, ADC, and so on,
% and uncovered 10 previously undisclosed vulnerabilities up to now.
% All the bugs were reported to vendors and patched in time.
% These vulnerabilities demonstrate \sys to be successful
% in fuzzing peripheral drivers and discovering relevant vulnerabilities.
% The implementation of the USB protocol on the STM32 SDK has all of these issues.
% We will conduct a comprehensive investigation into these vulnerabilities
% with documented protocol in the following.

\paragraph{Unsanitized Length Attributes in Descriptors.} 
% USB enumeration is the procedure by which the host seeks information 
% from the device and configures the communication channel with the device.
% The detailed procedure of the enumeration is shown in Figure~\ref{fig:usb_enum}.
% The USB protocol specifies the endpoint, 
% which is the hardware buffer in the device used for data transmission.
% On the host side, the endpoint is recognized as the software pipe.
% Vendors may manufacture USB devices with numerous optional endpoints
% but not less than one control endpoint, 
% known \textit{Endpoint0}.
% The host retrieves the device information or 
% sends the control commands through \textit{Endpoint0}.
% During enumeration, the host retrieves the device descriptor, 
% the configuration descriptor, the interface descriptor and the endpoint descriptor
% in succession as shown in steps \textcircled{5}, \textcircled{8},
% \textcircled{10}, \textcircled{12} of Figure~\ref{fig:usb_enum}.
Lots of bugs were caused by the failure to properly sanitize 
the length fields in different descriptors.
For example, the attribute \textit{bMaxPacketSize} of 
the device descriptor (\textcircled{6}) defines the allowed
maximum size of the packet transmitted from the host
to the device.
If this value is maliciously specified and not checked,
the driver in the MCU host would wrongly configure
a faulty pipe and allocate a buffer of unexpected size, causing a denial of service.
We found similar bugs in parsing the field \textit{wTotalLength} of the configuration
descriptor (\textcircled{9}), interface descriptor (\textcircled{11})
and endpoint descriptor (\textcircled{13}),
and the field \textit{wMaxPacketSize} of the endpoint descriptor \textcircled{13}.
The consequences of these bugs range from buffer overflow to denial of service.
The relevant CVEs in this category found by our tool include  
CVE-2021-34259, CVE-2021-34260, CVE-2021-34268, CVE-2021-38258 and CVE-2021-38260.

\paragraph{Missed Hardware Support Checking.} 
The field \textit{bmAttributes} in the configuration descriptor
indicates different kinds of power parameters of the configuration.
When the respective bit is set,
the USB driver attempts to activate the remote wake-up
feature in order to save power.
However, the driver does not check if the device implements this function, 
and the system will hang as a result of a failed request.
CVE-2021-34261 belongs to this category.
% The relevant CVEs in this category found by our tool include  
% {\color{red}{This bug is assigned with .}}
% https://github.com/STMicroelectronics/STM32CubeH7/issues/78

\paragraph{Illogical Endpoint Address.} 
The field \textit{bEndpointAddress} in the endpoint descriptor 
specifies the identification of the endpoint.
The host should generally establish an IN endpoint and
an OUT endpoint
for data reception and transmission.
The driver takes this for granted
and does not check the contents of the endpoint descriptor.
When a malformed endpoint descriptor
specifies the same direction,
the host may lose the IN pipe or OUT pipe,
leading to system hanging.
CVE-2021-34267  belongs to this category.
% https://github.com/STMicroelectronics/STM32CubeH7/issues/80

\paragraph{Unchecked Polling Interval.} 
The filed \textit{bInterval} in the endpoint descriptor
indicates the interval for polling endpoint data transmission.
It will be used to specify different polling intervals for
individual USB applications, such as audio streaming.
When this value is negative,
the polling logic will hang the device.
CVE-2021-34262 belongs to this category.
% {\color{red}{This bug is assigned with .}}
% The driver does not check if it is greater than zero 
% which value causes the polling logic to hang.
% https://github.com/STMicroelectronics/STM32CubeH7/issues/82

\section{Discussions}

\paragraph{Fuzzing Driver Code.}
Fuzzing peripheral drivers is different from fuzzing normal libraries because
peripheral drivers need to interact with the external physical world,
which brings lots of nondeterminism in fuzzing. 
% hardware in comparison with universal libraries.
% More specifically, the firmware needs to read from or write to memory-mapped registers
% to operate with peripherals.
In addition, peripheral behaviors are typically modeled
as a state machine, whose state transition is triggered
by many kinds of events such as interrupts and MMIO interactions.
Consequently, effectively fuzzing peripheral drivers needs profound domain knowledge,
% to target a particular state,
as we have demonstrated in fuzzing the USB enumeration implementation.
Because of this, the code harness for fuzzing a peripheral driver
is typically conducted case-by-case.
As a general guideline, we should find the program points
where the driver interacts with the hardware peripherals.
% via memory-mapped registers.
These are places that take inputs from untrusted sources and
should thus be tamed to read the testcases.
We plan to
fuzz more complex drivers in the future.

\paragraph{Applicability.}
Our approach relies on the ETM hardware feature and
therefore cannot be used to test chips without this feature.
Fortunately, based on our experience, ETM is very popular
among MCU chips.
The real problem is that 
only a small portion of development boards have physical pinouts to
interface with ETM, due to the additional cost in PCB design.
We argue this does not influence the applicability of our tool in
performing in-house testing or fuzzing vendor SDKs.
First, developers can easily assemble a PCB board
with ETM pinouts~\cite{pcb_design}.
Second, the result of fuzzing an SDK is typically
applicable to other SDKs of the same chip vendor.
Specifically, chip vendors tend to maintain a common set of device 
drivers for similar product lines. 
If a single development board has the ETM pinouts, 
the discovered bugs could apply to other chips. 
Taking STM32 MCUs as an example, initially, our tool 
found CVE-2021-34268 for the STM32H7 series chips since we only
have access to an STM32H7B3I-EVAL board which is ETM-enabled. 
However, the bug affects 
almost all the product lines of STM32 chips including 
STM32F4\footnote{\url{https://github.com/STMicroelectronics/STM32CubeF4/blob/2f3b26f16559f7af495727a98253067a31182cfc/Middlewares/ST/STM32_USB_Host_Library/Core/Src/usbh_ctlreq.c\#L355}}, 
STM32F1\footnote{\url{https://github.com/STMicroelectronics/STM32CubeF1/blob/f5aaa9b45492d70585ade1dac4d1e33d5531c171/Middlewares/ST/STM32_USB_Host_Library/Core/Src/usbh_ctlreq.c\#L360}}, etc.

\section{Related work}
\subsection{MCU Firmware Fuzzing}

Existing approaches for MCU firmware fuzzing can be classified into five categories as shown in Figure 4 of the work by Li et al~\cite{rehosting}.
Emulation is the most intuitive method. 
However, full emulation of various MCU in the market is impossible.
Existing QEMU-based solutions, 
such as {P$^{2}$IM}~\cite{P2IM}, DICE~\cite{dice}, Laelaps~\cite{laelaps}, PRETENDER~\cite{PRETENDER}, $\mu$Emu~\cite{uemu}, and Jetset~\cite{jetset} 
make a trade-off by utilizing a variety of methods, like machine-learning, symbolic execution and so on, to approximate the behavior of the peripherals in order to run the firmware successfully.
HALucinator~\cite{HALucinator} simplifies the process even further by abstracting and replacing the hardware interface without performing any actual peripheral functions.
All of them demonstrate good performance and cost-effectiveness. 
As with HALucinator, para-rehosting~\cite{rehosting} also provides a uniform platform by abstracting an MCU on the commodity hardware in order to
benefit from the off-the-shelf test suites such as AFL~\cite{afl} and AddressSanitizer~\cite{serebryany2012addresssanitizer} with the most powerful capability.
Furthermore, it can greatly improve the fuzzing performance.
Both techniques, however, are not able to delve into the driver layer. 
Although the peripheral forwarding mechanisms (\eg Avatar~\cite{avatar}) interact with hardware, 
they are primarily used to support the upper layer's fuzzing. 
% Besides, using the forwarding peripheral operations alone will cause the context loss problem. 
Also, full firmware synchronization between the emulator and the real device will incur additional overhead.

% when testing firmwares' upper layer

% HALucinator~\cite{HALucinator} and {P$^{2}$IM}~\cite{P2IM}, obtain a good result by abstracting and replacing the hardware interface when testing firmwares' upper layer, while the para-rehosting approach~\cite{rehosting} provides a uniform platform
% to port RTOS and HAL to the commodity hardware. Therefore, para-rehosting could benefit from the off-the-shelf test suites such as AFL~\cite{afl} and AddressSanitizer~\cite{serebryany2012addresssanitizer} with the most powerful capability. Furthermore, it can greatly improve the fuzzing performance.

% However, both techniques cannot delve into the driver layer. Although the peripheral forwarding mechanisms (\eg Avatar~\cite{avatar}) interact with hardware, they mainly support the fuzzing of the upper layer. Besides, using the forwarding peripheral operations alone will cause the context loss problem. Also, full firmware synchronization between the emulator and the real device will incur an additional overhead.

\subsection{Peripheral Vulnerability Detection}
USBFuzz~\cite{USSBFuzz} proposed a framework to apply fuzzing to the commodity OS USB driver
by using an emulated USB device in a virtualized kernel.
The emulated USB device feeds the testcases and breaks the hardware/software barrier.
However, this approach cannot be used in MCU firmware fuzzing, because
\begin{enumerate*}[label={\alph*)},itemjoin={{, }}, itemjoin*={{, and }}]
    \item unlike the kernel in a commodity OS host, MCU cannot run any existing fuzzer due to lack of MMU
    \item no out-of-box device emulators are available for MCU
    \item MCU peripherals are multifarious and heterogeneous.
\end{enumerate*}
Therefore, the MCU peripherals still have to be considered individually.

Facedancer~\cite{facedancer}, an early effort for fuzzing MCU USB drivers, proposed to use a board which responded with a variety of descriptors and disguised itself as different USB devices. Working as the man in the middle between the fuzzer and the target MCU, the board helps to feed the testcase and check the operating state. However, this dummy fuzzing approach is not effective in most cases, since it cannot obtain any feedback from the target to generate the bitmap and guide the testcase generation. Besides, the board supported only USB drivers.

FIRMUSB~\cite{hernandez2017firmusb} took a domain-specific approach for detecting USB vulnerabilities. To identify non-compliant behaviors, it contrasted the model generated from known USB databases and retrieved by symbolic execution. So, FIRMUSB could detect vulnerabilities of the binary firmware without running it on the real board. However, similar to FaceDance, this effort focuses on testing USB firmware only.
\section{Conclusion}
We propose \sys, a non-intrusive feedback-driven fuzzing platform
for MCU firmware,
particularly targeting the low-level peripheral drivers.
\sys decouples the execution engine from the original
AFL framework and uses the development board to execute
the testcase.
To enable communication between the execution engine and
the rest of AFL, we leverage the debug dongle commonly
available in the embedded system development environment.
To effectively obtain instruction trace,
we rely on the ETM and DWT hardware features. 
Finally, we propose using dynamic basic blocks to reduce the
overhead of decoding and analyzing the ETM data.
% \begin{enumerate*}[label={\alph*)},itemjoin={{, }}, itemjoin*={{, and }}, font={\color{red!50!black}\bfseries}]
%     \item collects the instruction trace
%     through debug dongle 
%     to recover the execution paths
%     \item provides online filters to focus only on 
%     interesting code snippet and improve the performance 
%     with the aid of another hardware feature DWT 
%     and an offline filter to solve the non-deterministic problem 
%     introduced by the interrupt
%     \item recover the execution path and represent each with dynBB
%     without disassembly to avoid the trouble of code obfuscation, 
%     compression problem and so on.
% \end{enumerate*}
% To verify the correctness and effectiveness,
We have evaluated our prototype implementation against SDKs from
the two popular MCU vendors:
NXP and STMicroelectronics.
The prototype has helped us find
13 zero-day bugs in the USB stack shipped with
the vendor SDKs,
and eight CVEs have been allocated.

 % {\color{red}{and gotten  }}.

\begin{acks}
We thank engineers from SEGGER Microcontroller
for their tireless technical support in using JLink SDK
and permitting us to distribute our work as an open-source project.
We thank
Dr. Kang Li from Baidu Security and Dr. Kyu Hyung Lee from the University of
Georgia for their insightful comments.
This work was supported
in part by NSF IIS-2014552, the Ripple University Blockchain Research Initiative and a grant from the University of Georgia Research Foundation, Inc.

\end{acks}

\bibliographystyle{ACM-Reference-Format}
\bibliography{bibs/symbolic,bibs/symbolic-execution,
bibs/dynamic-analysis,bibs/cortexm,
bibs/rtos, bibs/others}

% %%
% %% If your work has an appendix, this is the place to put it.
% \appendix









\end{document}